\documentclass[10pt, conference, letterpaper]{IEEEtran}
\usepackage{epsfig}
\usepackage{amsmath}
\usepackage{amssymb,amsthm}
\usepackage{amsbsy}
\usepackage{graphicx}
\usepackage{hyperref}
\usepackage{color}
\usepackage{ifthen}
\usepackage{epstopdf}
\usepackage{bbm}
\usepackage{caption}
\usepackage[labelfont=Large]{subfig}

\newtheorem{lemma}{Lemma}

\newtheorem{theorem}{Theorem}

\newtheorem{assumption}{Assumption}

\newboolean{showcomments}
\setboolean{showcomments}{true}

\graphicspath{ {./images/} }

\newcommand{\cp}{\mathcal{N}}

\newcommand{\scp}{\mathcal{S}}
\newcommand{\ncp}{\mathcal{O}}

\newcommand{\ra}{\rightarrow}

\newcommand{\indicator}[1]{\mathbbm{1}_{\{#1\}}}

\allowdisplaybreaks[4]

\IEEEoverridecommandlockouts

\begin{document}

\title{Sponsored data with ISP competition}

\author{
	\IEEEauthorblockN{Pooja Vyavahare \thanks{The  work  of  Pooja Vyavahare  was  supported  by  an  INSPIRE  Faculty Award of the DST, Government of India.}}
	\IEEEauthorblockA{\textit{EE, IIT Tirupati}}
	\and
	\IEEEauthorblockN{D. Manjunath and Jayakrishnan Nair \thanks{      
			The work of D Manjunath and J Nair was supported grants from CEFIPRA and DST. They are associated with the Bharti Centre for Communication at IIT Bombay.}}
	\IEEEauthorblockA{\textit{EE, IIT Bombay}} }
\maketitle
\begin{abstract}
  We analyze the effect of sponsored data platforms when Internet
  service providers (ISPs) compete for subscribers and content
  providers (CPs) compete for a share of the bandwidth usage by the
  customers. Our analytical model is of a full information,
  leader-follower game. ISPs lead and set prices for sponsorship. CPs
  then make the binary decision of sponsoring or not sponsoring their
  content on the ISPs. Lastly, based on both of these, users make a
  two-part decision of choosing the ISP to which they subscribe, and
  the amount of data to consume from each of the CPs through the
  chosen ISP. User consumption is determined by a utility maximization
  framework, the sponsorship decision is determined by a
  non-cooperative game between the CPs, and the ISPs set their prices
  to maximize their profit in response to the prices set by the
  competing ISP. We analyze the pricing dynamics of the prices set by
  the ISPs, the sponsorship decisions that the CPs make and the market
  structure therein, and the surpluses of the ISPs, CPs, and users.

  This is the first analysis of the effect sponsored data platforms in
  the presence of ISP competition. We show that inter-ISP competition
  does not inhibit ISPs from extracting a significant fraction of the
  CP surplus. Moreover, the ISPs often have an incentive to
  significantly skew the CP marketplace in favor of the most
  profitable CP.

\end{abstract}

\maketitle

\section{Introduction}
\label{sec:intro}

Market segmentation and discriminatory pricing are well known
techniques \cite{Varian10,Tirole88} that ISPs can use to increase
revenues. A combination of inter-ISP competition and market
expectations have rendered such schemes to be not so prevalent on the
user side. Regulatory issues have also prevented the use of many smart
data pricing schemes. However, sponsored data or zero-rating is a
price discrimination technique that is being introduced by ISPs in
many markets as a consumer friendly innovation and is gaining
increased adaptation. In this scheme, the content provider (CP) pays
the ISP charges for its content that is consumed by the users while
the users do not pay the ISP charges for the same.

Regulatory response to sponsored data, or zero-rating, has been
varied. In many countries, it is deemed to violate net neutrality
regulations and is hence banned, e.g., Canada, Brazil, India, Chile,
Sweden, Hungary. In many other countries it is allowed alongside net
neutrality regulations that disallow discriminatory QoS schemes, e.g.,
USA, UK, Netherlands, Germany \cite{Sandvine16b}. In fact, BEREC
guidelines stipulate a case by case analysis when zero rating is a
purely pricing practice, and leaves it to the national regulatory
authorities. Wherever allowed, it is expected that such schemes will
become more prevalent and many companies are making plans to enter
this 23 billion dollar
market\footnote{\path{https://www.mobilemarketer.com/ex/mobilemarketer/cms/news/research/20919.html}}.
AT\&T is revamping its Data Perks program to offer free DirecTV and
other video
services\footnote{\path{https://www.theverge.com/2018/2/20/17032550/att-prepaid-plans-sponsored-data}},
Verizon is offering AOL Gameday and Hearst magazines via its FreeBee
program\footnote{\path{https://www.theverge.com/2016/1/19/10789522/verizon-freebee-sponsored-data-net-neutrality}}
and T-Mobile has been offering free music on BingeOn. There are also
third party providers for such services, e.g.,
Aquto\footnote{\texttt{http://www.aquto.com/}}.

In this paper we study the effect of such services on the content
provider market and on the surpluses of various stakeholders.

\subsection{Previous work}
The economics of discrimination and its effect on market structures,
on investment incentives, and on stakeholder surpluses have been
widely studied. In the provisioning of Internet service, one version
of discrimination is called QoS discrimination. This is effected
either by providing fast lanes for preferred CPs or by giving them
transmission priorities or a combination of the two. The effect of QoS
discrimination is analyzed in, e.g.,
\cite{Musacchio09,Choi10,Economides12a}.  With QoS discrimination, the
improved quality of experience drives users toward the preferred
CPs. Our interest in this paper is in price discrimination effected
through a sponsorship program or a zero-rating platform. In this
scheme, the content of the sponsoring or zero-rated CPs is free to the
user while the user pays for the content from the non sponsoring
CPs. Here, cheaper prices drives users towards sponsored or zero-rated
content. (In the rest of the paper we will use the terms sponsored
data and zero-rating interchangeably.)  Examples of work that address
price discrimination are
\cite{Andrews13a,Lotfi15,Ma14,Joe-Wang15,Zhang15,Phalak18}.  In
\cite{Andrews13a,Lotfi15}, one ISP and one CP interact in a
Stackelberg game. Two CPs and one ISP are considered in
\cite{Ma14,Joe-Wang15,Zhang15,Phalak18,Somogyi17,Jullien18}. These
papers differ in the interaction between the agents, the consumption
model for the users, and the manner in which sponsorship is effected.
However, the key conclusion in all of them is qualitatively
similar---as the revenue rate of the CPs increase, the ISP can achieve
higher profits than in the case where sponsorship is not
allowed. Further, CPs with lower revenue rate possibly lose more on
their surplus either due to sponsorship costs, or due to competition
with free content.  This can make them become less profitable in the
short term and potentially nonviable in the long term.

All of the preceding work considered only one ISP and this begs the
natural question: Would ISP competition reduce the ability of the ISPs
to extract the CP surplus? Specifically, would the ISPs be as powerful
as the models that use only one ISP indicate. Surprisingly, the answer
is in the affirmative, albeit with some qualifications.

We mention here that the only prior work that we are aware of that
considers zero-rating with ISP-competition is \cite{Maille17}. This is
a purely numerical study, where the strategic interaction between the
competing ISPs and the resulting equilibria are not considered.

\subsection{Preview}
In the next section we set up the notation and the model for the
leader-follower game involving two ISPs as leaders, two CPs following
the ISPs, and a continuum of users following the CPs. We begin by
describing the user behavior for a given set of ISP prices and CP
sponsorships. This is then used to determine the market share of the
ISPs. We then describe the how the CPs make the sponsorship decision
for a given set of prices from the ISPs. Finally, the determination of
the prices by the ISPs is also detailed.

In Section~\ref{sec:opt_q1} we derive the best response strategies of
one ISP for a given set of prices and the sponsorship configuration of
the CPs on the competing ISP. The key contribution in this section is
that
\begin{itemize}
\item there is a threshold on CP profitability beyond which the ISP
  will price its data sponsorship service such that at least one of
  the CPs will sponsor its data, and
\item at least one of the CPs has less surplus than it would have had
  if the ISP did not operate a data sponsorship program.
\end{itemize}

In Section~\ref{sec:examples}, we use the results of the previous
section to have the ISPs sequentially determine their best response
prices in response to the sponsorship configuration on the competing
ISP, in a t\^atonnement-like iterative process. The following are the
key contributions in this section.
\begin{itemize}
\item For a wide range of parameter sets, we find that numerically,
  the iterative process does converge to an equilibrium rather
  quickly. Further, at equilibrium the CPs choose the same
  configuration on both the ISPs.
\item In some cases, the equilibrium configuration is the same as that
  in the case where each ISP acts as a monopoly. As a result, the ISPs
  are unaffected by inter-ISP competition, and both ISPs are able to
  extract a fraction of the CP-side surplus. At least one CP, and
  sometimes both CPs, end up worse off in the process, compared to the
  scenario where data sponsorship is not permitted.
\item In some cases, inter-ISP competition results in a
  \emph{prisoner's dilemma}, causing both ISPs to induce a sub-optimal
  sponsorship configuration. However, this does not necessarily result
  in a benefit for CPs. At least one CP, and sometimes both CPs, still
  end up worse off, compared to the scenario where data sponsorship is
  not permitted.
\end{itemize}

Finally, in Section~\ref{sec:asymmetric} we consider asymmetric
stickiness of the customers and see that this does not qualitatively
change our conclusions above.

We conclude with a discussion and some policy prescriptions in
Section~\ref{sec:discuss}. The key input to policy planners from the
preceding is that although ISP and CP competition can provide price
stability, data sponsorship practices enable ISPs to extract a
substantial portion of CP surplus---importantly, this ability is not
diminished by inter-ISP competition. The resulting asymmetry of
benefits drives smaller CPs towards significantly lower profitability
and possibly exiting the market. Thus, while data sponsorship may
provide improved surplus to users in the short run, it can also
diminish competition in the CP marketplace.

\section{Model and preliminaries}
\label{sec:model}	

We consider two competing ISPs and two competing CPs. Each ISP
operates a zero-rating platform, and CPs have the option of sponsoring
their content by joining the zero-rating platform of one or both
ISPs. ISP~$j$ ($j \in \{1,2\}$) charges $p_j$ dollars per unit of data
to its subscribers and a sponsoring charge of $q_j$ dollars per unit
of data on CPs that zero-rate their content.\footnote{Such usage-based
  pricing is prevalent in the mobile Internet space \cite{Ha12}.}  CPs
derive their revenue via advertisements; CP~$i$ ($i \in \{1,2\}$)
makes a revenue of $a_i$ dollars per unit of data consumed by
users. Users subscribe to exactly one of the two ISPs and consume
content of the CPs through that ISP. Further, the volume of user
consumption is determined by the ISP charges and the utility obtained.

We capture the strategic interaction between the users, CPs, and ISPs
via a three-tier leader follower model:
\begin{enumerate}
\item ISPs `lead' by setting sponsorship charges. For simplicity, we
  assume that user charges are equal, i.e., $p_1 = p_2 = p,$ and are
  exogenously determined. 
  \footnote{Indeed, in many markets, user expectations and inter-ISP
    competition have driven user-side pricing to be flat across
    providers.}
\item CPs respond to sponsorship charges by making the binary decision
  of whether or not to sponsor their content on each ISP.
\item Finally, the user base responds to the actions of the CPs by
  determining the fraction of subscribers of each ISP. Moreover,
  subscribers of each ISP determine their consumption of each CP's
  content.
\end{enumerate}
In the following, we describe in detail our behavior model of the user
base, followed by our models for the behavior of the CPs and the
ISPs. Proofs of the results stated in this
  section can be found in Appendix~\ref{sec:app_NE-conditions}.

\subsection{User behavior}
\label{sec:user_model}

We begin by describing the consumption profile of users of ISP~$j$
($j \in \{1,2\}$), and subsequently describe how the market split
across ISPs is determined

\subsubsection*{Behavior of users of ISP~$j$} Let $\cp =\{1,2\}$
denote the set of CPs.
The set of sponsoring CPs on ISP $j$ is denoted by $\scp_j$ and
$\ncp_j = \cp \setminus \scp_j$ denotes the set of non-sponsoring CPs
on ISP $j.$ We denote the configurations $\scp_j = \emptyset,$
$\scp_j = \{1\},$ $\scp_j=\{2\},$ and $\scp_j =\{1,2\}$ by NN, SN, NS
and SS respectively.

We assume that users derive a utility of $\psi_i(\theta)$ from
consuming $\theta$ bytes of data from CP~$i$ within a billing
cycle. Here, $\psi_i(\cdot): \mathbb{R}_+ \rightarrow \mathbb{R}_+$ is
a continuously differentiable, concave and strictly increasing
function. We further assume that each user has a `capacity to consume'
$c$ bytes, which is the maximum amount of data (across both CPs) a
user can consume in a billing cycle. Let $\theta_{i,j}$ denote the
consumption of CP~$i$ content by users of ISP~$j.$ Thus, we take
$\theta_j = (\theta_{i,j}, i \in \cp)$ to be the unique solution of
the following optimization.
\begin{equation}
  \label{eq:user-opt}
  \begin{array}{rl}
    \displaystyle \max_{z = (z_1,z_2)} & \sum_{i \in \cp} \psi_{i}(z_i) - p \sum_{i
                           \in \ncp_j} z_i \\
    \displaystyle \text{s.t. } & \sum_{i \in \cp} z_i \leq c,\quad z \geq 0
  \end{array}
\end{equation}
The first term in the objective function above is the utility derived
from content consumption, and the second term is the price paid by the
user to ISP~$j$ for the consumption of non-sponsored content. Since
$p$ is assumed to be determined exogenously, it follows that the
solution of the above optimization depends only on the sponsorship
configuration $M_j \in \{$NN,SN,NS,SS$\}$ on ISP~$j.$ We sometimes
write the solution of~\eqref{eq:user-opt} as $\theta^{M_j} =
(\theta_{i}^{M_j}, i \in \cp)$ to emphasize this dependence.
We denote the optimal value of \eqref{eq:user-opt} by $u^{M_j}.$

Throughout this paper, we make the assumption that the two CPs are
substitutable, i.e., $\psi_1(\cdot) = \psi_2(\cdot) = \psi(\cdot);$
this simplifies notation and also enables us to highlight the impact
of zero-rating in \emph{skewing} the user consumption
profile.\footnote{We do however explicitly capture asymmetry in the CP
  revenue rates $a_j.$ Indeed, different CPs that offer comparable
  services may differ in their ability generate ad revenue.} However,
several of our results (including those stated in
Section~\ref{sec:opt_q1}, along with Theorems~\ref{thm:symmetric}
and~\ref{thm:isp_eq_a_large_rho_small} in Section~\ref{sec:examples})
generalize for $\psi_1(\cdot) \neq \psi_2(\cdot).$
Under the CP-substitutability assumption, it is easy to see that the
surpluses of users of ISP~$j$ under different sponsorship configurations
are sorted as follows.
\begin{lemma}
  \label{lemma:user_utility_sort}
  $u^{SS} \geq u^{SN} = u^{NS} \geq u^{NN}.$
\end{lemma}
Finally, we note the following consequence of the above consumption
model.
\begin{lemma}
  \label{lemma:theta_ordering}
  For any sponsorship configuration $M,$ and for $i \in \cp,$
  $\theta^{SN}_2 = \theta^{NS}_1 \leq \theta^M_i.$
\end{lemma}

Having described the content consumption profile of users of each ISP,
we now describe how the user base gets divided across the ISPs. Toward
that, we assume that the users can change their ISP subscription at
any time.

\subsubsection*{Market split across ISPs}

We model the distribution of users between ISPs using the
\textit{Hotelling model} \cite{Hotelling1990}. Let $x = x_{M_1}^{M_2}$
denote the fraction of the user base subscribed to ISP~1. Under the
Hotelling model, $x_{M_1}^{M_2}$ is the solution of the equation
\begin{align}
  \label{eq:Hotelling}
  u^{M_1} - tx_{M_1}^{M_2} = u^{M_2} - t(1-x_{M_1}^{M_2}),
\end{align}
where $t > 0$ is a parameter of the model. This equation may
interpreted as follows: We imagine the users as being distributed
uniformly over the unit interval $[0,1].$ ISP~1 is located at the left
end-point of this interval, and ISP~2 is located at the right
end-point. A user at position $x \in [0,1]$ incurs a (virtual)
transportation cost of $tx$ to connect to ISP~1, and a (virtual)
transportation cost $t(1-x)$ to connect to ISP~2. Since each
(non-atomic) user connects to the ISP that provides the higher payoff
(surplus minus transportation cost), the market split is determined by
\eqref{eq:Hotelling}. Note that the transportation cost captures the
inherent \emph{stickiness} of users to a certain ISP; users located in
the left (respectively, right) half of the interval have an inherent
preference for ISP~1 (respectively, ISP~2).\footnote{In practice, user
  stickiness may result from many considerations like inertia, high
  lead time to switch ISPs, and familiarity with the features and
  services offered by one's present ISP.}  Moreover, a higher value of
$t$ implies increased user stickiness. To ensure a meaningful solution
to \eqref{eq:Hotelling}, we assume that $t > u^{SS} - u^{NN}.$ It then
follows that the market share of ISP~1 is given by
\begin{equation}
  \label{eq:Hotelling_2}
  x_{M_1}^{M_2} = \frac{u^{M_1}-u^{M_2}+t}{2t}.
\end{equation}

Note that the Hotelling model has been extensively used in many
similar situations, including in the modeling of ISPs, e.g.,
\cite{Choi10}. Further, a generalization is considered in
Section~\ref{sec:asymmetric} where we will assume that the stickiness
of the users is not symmetric, i.e., the $t$ is different for
different ISPs. 

We conclude by collecting some immediate consequences of the Hotelling
model.

\begin{lemma}
	\label{lemma:market_share_order}
  For any given sponsorship configuration $M_2$ on ISP2, the market
  market share of ISP1 under different sponsorship configurations are
  related as follows:
  \[x_{NN}^{M_2} \leq x_{SN}^{M_2} = x_{NS}^{M_2} \leq x_{SS}^{M_2}. \]
\end{lemma}
This lemma is an immediate consequence of
Lemma~\ref{lemma:user_utility_sort} and~\eqref{eq:Hotelling_2}.

\begin{lemma}
\label{lm:t_infinity_market}
As $t \ra \infty,$ for any given sponsorship configurations $M_1$ and
$M_2$ on ISPs~1 and~2, $x_{M_1}^{M_2} \rightarrow 0.5.$
\end{lemma}
The above lemma, which is a direct consequence of
\eqref{eq:Hotelling_2}, states that as user stickiness grows, the
market shares of the ISPs become insensitive to their sponsorship
configurations and approach a symmetric market split.
In other words, as $t \ra \infty,$ the churn of users between ISPs
diminishes, and each ISP can be thought of as a \emph{monopoly}.

\subsection{CP behavior}
\label{sec:cp_behaviour_isp1}

In this subsection, we describe our model of CP behavior. Recall that
in our leader-follower model, CPs lead the users and follow the ISPs,
i.e., they decide whether on not to sponsor their content on ISPs~1
and~2 based on sponsorship charges announced by the ISPs, knowing
\emph{ex-ante} that the user base will respond to their actions based
on the model presented in Section~\ref{sec:user_model}. Since each CP
seeks to maximize its own profit, it is natural to capture the outcome
of their interaction as a Nash equilibrium.

Note that in general, each CP may choose to either sponsor or not
sponsor its content on each ISP. This means that there are four
possible actions per CP, and sixteen possible sponsorship
configurations in all. To avoid the resulting analytical (and
notational) complexity, we make the following simplifying assumption.
\begin{assumption}
  \label{assumption:CP}
  The CPs can only reconsider their sponsorship decision on a single
  ISP at a time.
\end{assumption}
Assumption~\ref{assumption:CP} is natural if there is a contractually
binding period associated with the decision to sponsor one's content
on an ISP, say ISP~1, with the opportunity to form (or renew) a
sponsorship contract with ISP~1 arising periodically and out of sync
with similar opportunities to sponsor on ISP~2.

Under Assumption~\ref{assumption:CP}, it is meaningful to ask the
question: \emph{Given a sponsorship configuration $M_2 \in
  \{$NN,SN,NS,SS$\}$ on ISP~$2,$ when is $M_1 \in \{$NN,SN,NS,SS$\}$ a
  Nash equilibrium sponsorship configuration on ISP~1?}  In the
remainder of this section, we address this question.

Consider an arbitrary sponsorship $M_2$ configuration on ISP~2. If
CP~$1$ chooses to sponsor its content on ISP~1, its surplus is given
by
\begin{align*}
  x (a_1 - q_1) \theta_{1,1} &+ (1-x) \bigl[ (a_1 - q_2) \theta_{1,2}  \indicator{M_2
  \in \{SS,SN\}} \\
  &+ a_1 \theta_{1,2} \indicator{M_2 \in \{NS,NN\}} \bigr].
\end{align*}
The first term above captures the surpluses from ISP~1 (revenue from
ISP~1 users minus the sponsorship charge paid to ISP~1). Note that
this term contains the market share of ISP~1 as a factor; also recall
that we take the `volume' of the user base to be unity. The second
term captures the surplus of CP~1 from ISP~2. Similarly, if CP~2
chooses not to sponsor its content on ISP~1, its surplus is given by
\begin{align*}
  x a_1 \theta_{1,1} &+ (1-x) \bigl[ (a_1 - q_2) \theta_{1,2}  \indicator{M_2
  \in \{SS,SN\}} \\
  &+ a_1 \theta_{1,2} \indicator{M_2 \in \{NS,NN\}} \bigr].
\end{align*}
It is important to note that in the above equations, $x,$
$\theta_{1,1}$ and $\theta_{1,2}$ depend on the actions of both
CPs. The
  conditions for the different sponsorship configurations on ISP~1 to
  be a Nash equilibrium are derived in
  Appendix~\ref{sec:app_NE-conditions}.

\subsection{ISP behavior}
\label{sec:ISPbehavior}

We now describe our model for ISP behavior. ISPs derive their revenue
from two sources: from users (subscribers) for the consumption of
non-sponsored content, and from CPs for the consumption of sponsored
content. Thus, the surplus of ISP~1 is given by $x \left[ \sum_{i \in
    \scp_1} q_i \theta_{i,1} +\sum_{i \in \ncp_1} p \theta_{i,1}
  \right],$ whereas that of ISP~2 is given by $(1-x) \left[ \sum_{i
    \in \scp_2} q_i \theta_{i,2} +\sum_{i \in \ncp_2} p \theta_{i,2}
  \right].$

The ISPs, being leaders of the three-tier leader-follower interaction,
set sponsorship prices as to induce the most profitable Nash
equilibrium among CPs on their zero-rating platform. Specifically, we
assume that given a sponsorship configuration on, say ISP~2, when
ISP~1 sets the sponsorship price on its zero-rating platform, the most
profitable (for ISP~1) Nash equilibrium between the CPs on its
platform emerges.\footnote{Note that if the action of an ISP
    allows for multiple Nash equilibria between the CPs (as per
    Lemma~\ref{lm:equilibrium_conditions}), we assume the ISP is able
    to induce the most profitable equilibrium. This is a standard
    approach for handling non-unique follower equilibria in
    leader-follower interactions \cite{kulkarni2015existence}.} Note
that in our model, the impact of the action of any ISP depends on the
prevailing sponsorship configuration on the other. In other words, the
interaction between the ISPs has \emph{memory}.

We define the tuple $(q_1,M_1,q_2,M_2)$ to be a system equilibrium if,
for $j \in \{1,2\},$\footnote{For any ISP~$j,$ we use the label $-j$ to
  refer to the other ISP.}
\begin{enumerate}
\item Given sponsorship configuration $M_{-j}$ on ISP~$-j,$ $M_j$ is
  the most profitable Nash equilibrium (among the CPs) for ISP~$j$
  under action $q_j.$
\item Given sponsorship configuration $M_{-j}$ on ISP~$-j,$ the
  surplus of ISP~$j$ is maximized under action $q_j.$
\end{enumerate}

Note that under a system equilibrium, neither ISP has the incentive to
unilaterally deviate from its action. Moreover, neither CP has the
unilateral incentive to deviate from its sponsorship decision on one
ISP given the prevailing sponsorship configuration on the other
ISP.\footnote{This is a weak notion of equilibrium, in the sense that
  it does not guarantee that CPs do not have the incentive to reverse
  their sponsoring decisions on \emph{both} ISPs. However, we prove in
  Section~\ref{sec:examples} that the system equilibria we observe do
  indeed possess this guarantee; see
  Theorems~\ref{thm:isp_eq_a_large_rho_small}
  and~\ref{thm:isp_eq_a_large_rho_large}.}

In Section~\ref{sec:opt_q1}, we explore the optimal response of ISP~1
given a prevailing sponsorship configuration on ISP~2. Then, in
Section~\ref{sec:examples}, we investigate system equilibria by
simulating best response dynamics between the ISPs.

\section{ISP's best response strategy}
\label{sec:opt_q1}

In this section, we assume a fixed sponsorship configuration $M_2$ on
ISP~2, and analyze the optimal strategy for ISP~1. This optimal
strategy involves setting the sponsorship charge $q_1$ so as to induce
the most profitable Nash equilibrium between the CPs on its
zero-rating platform. We also study the impact of ISP~1's optimal
strategy on the surplus of both ISPs, both CPs, and the user base.

The analysis of this section sheds light on the behavior we might
expect from an ISP in a competitive marketplace. Indeed, the case
$M_2$ = NN can also be thought as capturing the scenario where one ISP
(ISP~1) operates a zero-rating platform, whereas the other (ISP~2)
does not. Such a situation can happen when the competing ISP is slow
to act; e.g., Sprint announced its zero rating service much later than
its competitors. Moreover, the analysis of this section captures one
step in the alternating best response dynamics we consider in the
following section, providing insights into the observed system
equilibria.

For notational simplicity, throughout this section, we take $M_2$ =
NN. Our results easily generalize to arbitrary $M_2.$
As in \cite{Phalak18}, we find it instructive to analyze ISP~1's
optimal strategy in the scaling regime of \emph{growing CP revenue
  rates}. After all, it is when CP revenue rates are large that ISPs
have the incentive to offer zero-rating opportunities, so as to
extract some of the CP-side surplus. Specifically, we consider
$(a_1,a_2) = (a,\rho a),$ where $a > 0$ is a scaling parameter and
$\rho \in (0,1)$ is fixed.
When $\rho$ is small, this corresponds to the scenario where CP~1 has
a considerably greater ability to monetize its content than CP~2,
although their services are comparable from a user standpoint. As we
shall see in this section and the next, the outcomes corresponding to
this case differ considerably from the outcomes when $\rho \approx 1,$
i.e., the CPs are comparable in their ability to monetize their
content.The proofs of the results claimed in this section can be found in the
  appendix.

The following theorem sheds light on the sponsorship configurations
induced by ISP~1 in the regime of growing CP revenue rates.

 \begin{theorem}
  \label{thm:opt_ISP1_market_str}
  \textbf{[ISP~1's profit maximizing strategy]} Let
  $(a_1,a_2) = (a,\rho a)$ where $a>0,$ and fixed $\rho\in (0,1).$
  There exists a threshold $a_s>0$ such that
  \begin{enumerate}
  \item For $a\leq a_s,$ ISP~1 enforces an NN equilibrium.
  \item For $a\geq a_s,$ ISP~1 enforces an SN/SS
    equilibrium.
    
  \end{enumerate}
  \label{thm:isp_q1}
 \end{theorem}

Theorem~\ref{thm:opt_ISP1_market_str} shows that when the revenue
rates of both CPs are small, ISP~1 favors an NN configuration, since
charging users is more profitable than charging the CPs. When the
revenue rates cross a certain threshold, ISP~1 induces an SN/SS
equilibrium depending on the values of $a$ and $\rho.$ Specifically,
if $\rho$ is small, i.e., CP~1 has a considerably higher revenue rate
than CP~2, then ISP~1 favors an SN configuration.
Indeed, in this case, it is in the interest of ISP~1 to skew user-side
consumption in favor of CP~1, thanks to the greater potential for
sponsorship revenue from CP~1 compared to CP~2. On the other hand,
when $\rho \approx 1,$ ISP~1 favors an SS configuration for large
enough $a.$
The proof of
  Theorem~\ref{thm:opt_ISP1_market_str} (see
  Appendix~\ref{sec:opt_ISP1_market_str}) also highlights the precise
  conditions for different optimal strategies for ISP~1.

Next, we note that the threshold on CP revenue rates for sponsorship
to be profitable for ISP~1 shrinks as user stickiness decreases. This
is because when user stickiness is small (i.e., $t$ is small), ISP~1
sees a sharp growth in its subscriber base once sponsorship kicks~in.

\begin{lemma}
  \label{lemma:opt_ISP1_threshold}
  Let $(a_1,a_2) = (a,\rho a)$ where $a>0,$ and fixed $\rho\in (0,1).$
  The sponsorship threshold $a_S$ defined in
  Theorem~\ref{thm:opt_ISP1_market_str} is an increasing function of
  $t.$
\end{lemma}

Our next result highlights the benefit to ISP~1 from zero-rating.

\begin{lemma}
  \label{thm:thm:opt_ISP1_ISP1rev}
  \textbf{[ISP~1 surplus]}
  Let $(a_1,a_2) = (a,\rho a)$ where $a>0,$ and fixed $\rho\in (0,1).$
  Under the optimal strategy for ISP1 (given by
  Theorem~\ref{thm:isp_q1}), the profit $r_I(a)$ of ISP~1 varies with
  $a$ as follows.
  \begin{enumerate}
  \item $r_I(a)$ is constant over $a\leq a_s.$
  \item For $a>a_s,$ $r_I(a)$ is a strictly increasing, super linear
    function of $a,$ i.e., there exist constants $\nu > 0$ and
    $\kappa$ such that $r_I(a) \geq \nu a + \kappa$ for $a>a_s.$
  \end{enumerate}
\end{lemma}

Note that for $a > a_s,$ ISP~1 profit grows at least linearly in $a,$
implying that ISP~1 is able to extract a fraction of the CP revenues
by optimally setting the sponsorship charge on its zero-rating
platform.

Next, we turn to CP-side surplus. As the following lemma shows, the
zero-rating platform leaves at least one CP worse off.
\begin{lemma} \textbf{[CP surplus]}
  Under the optimal strategy for ISP1 (given by
  Theorem~\ref{thm:isp_q1}), the following statements hold.
  \begin{enumerate}
  \item When ISP~1 induces an SN equilibrium, CP~1 makes the same
    profit as it would make under an NN configuration (or
    equivalently, without the zero rating platform on ISP~1). On the
    other hand, CP~2 makes a profit less than or equal to that it
    would make under an NN configuration.
  \item Under an SS equilibrium, at least one of the CPs makes a
    profit less than or equal to that it would make under an NN
    configuration.
  \end{enumerate}
  \label{thm:cp_rev_only_p}
\end{lemma}

Finally, we note that since zero-rating increases user surplus (see
Lemma~\ref{lemma:user_utility_sort}), it is clear that the surplus of
subscribers of ISP~1, and also the aggregate surplus of the user base,
is increased for $a > a_S.$

To summarize, the results in this section show that so long as the CP
revenue rates are large enough, ISP~1 can set the sponsorship charges
on its zero-rating platform so as to extract a considerable fraction
of CP-side surplus, leaving one or both the CPs worse off. Moreover,
ISP~1 also benefits from the growth of its subscriber base that results
from the increased utility afforded to its users from sponsorship.

\begin{figure}
	\begin{center}
		\resizebox{\linewidth}{!}{
			\includegraphics{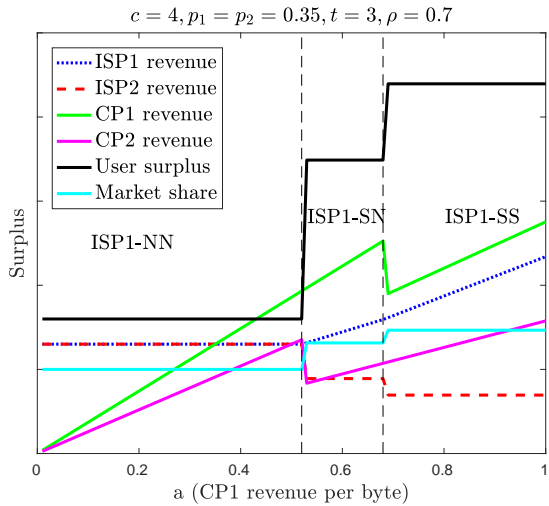}
		}
		\caption{Surplus of various entities in the system as a function of $a$ when ISP2 is in NN configuration.}
		\label{fig:surplus_is2_fix_t3_rho7}
	\end{center}
\end{figure}

We illustrate the results of this section with a numerical
  example. Specifically, we take $\psi(\theta)=\log(1+\theta),$
  $p = 0.35,$ $c = 4,$ $t = 3$ and $\rho = 0.7.$
  Figure~\ref{fig:surplus_is2_fix_t3_rho7} shows the surplus of the
  ISPs and CPs as a function of $a.$ Note that for intermediate values
  of $a,$ ISP~1 induces an SN equilibrium, and for larger values of
  $a,$ ISP~1 induces an SS equilibrium. Moreover, in the SS
  configuration, both CPs are worse off compared to the NN
  configuration (the profit under NN can be visualized by extending
  the linear growth in profit for small $a$). This can be thought of
  as a \emph{prisoner's dilemma} between the CPs. Finally, note that
  ISP~2 surplus shrinks once ISP~1 induces an SN configuration, and
  shrinks further when ISP~1 induces an SS configuration.

While the present section only considers the strategic behavior of a
single ISP, in the following section, we seek to capture the strategic
interaction between the ISPs.

\section{Equilibria of best response dynamics}
\label{sec:examples}

The goal of this section is to study the strategic interaction between
the ISPs, each ISP seeking to maximize its own profit. Since a
characterization of the system equilibria associated with the
three-tier interaction between the ISPs, the CPs, and the users is not
analytically feasible (except in two limiting regimes; see
Theorems~\ref{thm:isp_eq_a_large_rho_small} and
\ref{thm:isp_eq_a_large_rho_large}), we explore the system equilibria
obtained by simulating alternating best response dynamics between the
ISPs, i.e., the ISPs alternatively play their optimal response to the
prevailing sponsorship configuration on the other ISP. (The results of
the previous section shed light on this optimal response.) These
dynamics capture a myopic interaction between competing ISPs. Note
that an equilibrium of these dynamics, i.e., a configuration where
neither ISP adapts its action, is also a system equilibrium as defined
in Section~\ref{sec:ISPbehavior}. In this section, we analyze the
properties of these equilibria (when they exist), highlighting the
resulting sponsorship configurations, and also the surplus of the
various parties. 

Our numerical experiments yield two interesting observations:
\begin{itemize}
\item The alternating best response dynamics either converge quickly
  (in 5 to 8 rounds) or (in some cases) oscillate indefinitely.
  \item When the dynamics do converge, the equilibrium is
  \emph{symmetric}, i.e., of the form $(q,M,q,M).$
\end{itemize}
This last observation leads us to analyze the implications of
symmetric system equilibria:
\begin{theorem}
  \label{thm:symmetric}
  Under any symmetric system equilibrium of the form $(q,M,q,M),$ the
  following holds.
  \begin{enumerate}
  \item If $M \in \{SN,NS\},$ then the CP that sponsors on both ISPs
    makes the same profit as it would if zero-rating were not
    permitted. On the other hand, the CP that does not sponsor on both
    ISPs makes a profit less than or equal to that it would if
    zero-rating were not permitted.
  \item If $M = SS,$ at least one of the CPs makes a profit less than
    or equal to that it would make if zero-rating were not permitted.
  \end{enumerate}
\end{theorem}
Theorem~\ref{thm:symmetric} highlights that under any symmetric system
equilibrium, at least one of the CPs is worse off, compared to the
case where zero-rating is not permitted. In the absence of inter-ISP
competition, a similar observation was made in \cite{Phalak18};
Theorem~\ref{thm:symmetric} highlights that \emph{competition at the
  ISP level does not necessarily translate to improved surplus at the
  CP level}. The proof of Theorem~\ref{thm:symmetric} is presented
  in Appendix~\ref{sec:sym_equilibrium_cp_proof}.

We now report the results of our numerical experiments. Throughout, we
use $\psi(\theta) = \log(1 + \theta).$ We set the initial
configuration on ISP~2 to be NN, and allow ISP~1 to play first
(although we observe that the limiting behavior of the dynamics is
robust to the initial condition).
 
\subsection{Equilibrium sponsorship configurations}
\label{sec:symmetric_market}

\begin{figure*}
  \begin{center}
    \resizebox{\linewidth}{!}{
      \subfloat[]{
	\includegraphics{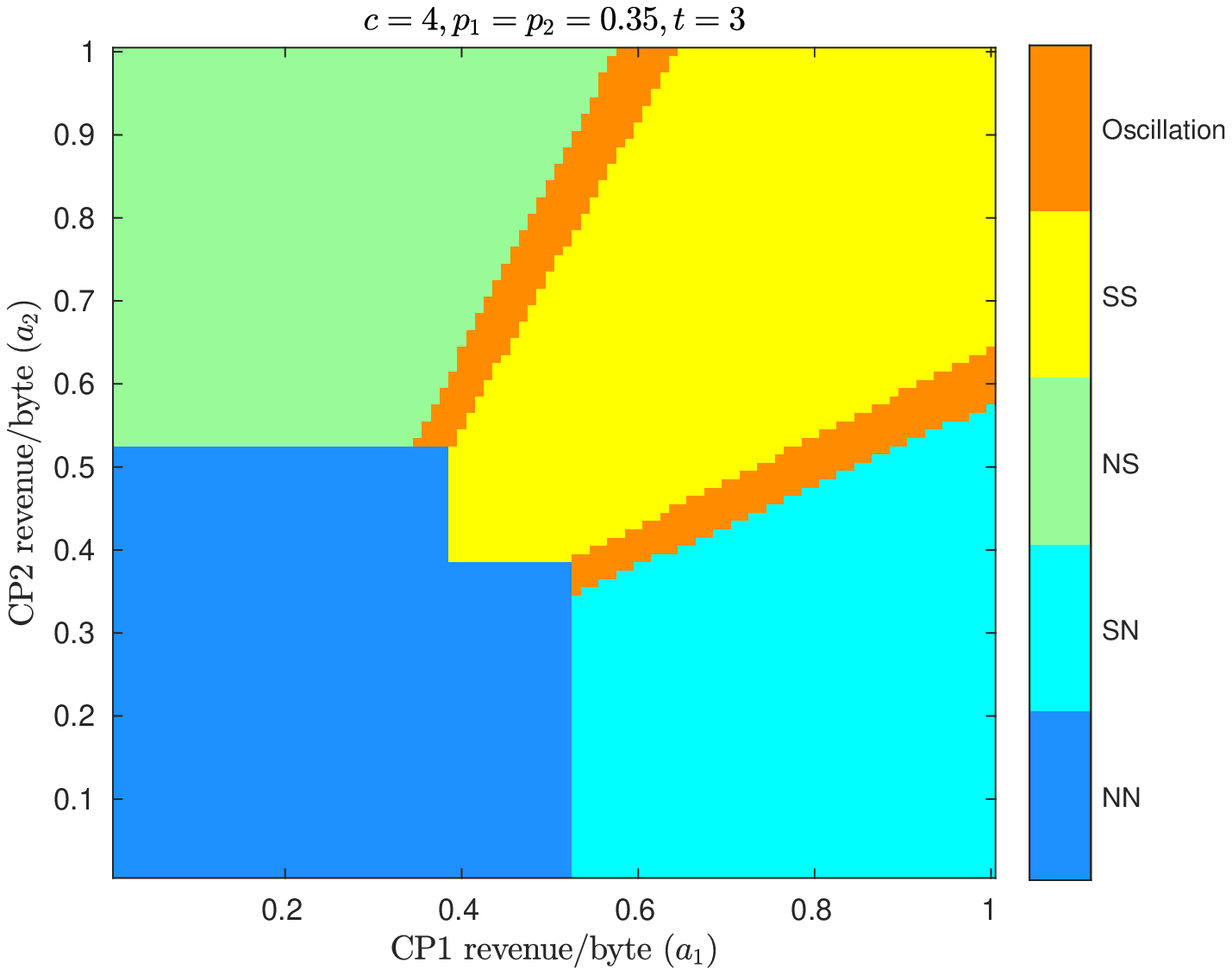}
      }
      \subfloat[]{
	\includegraphics{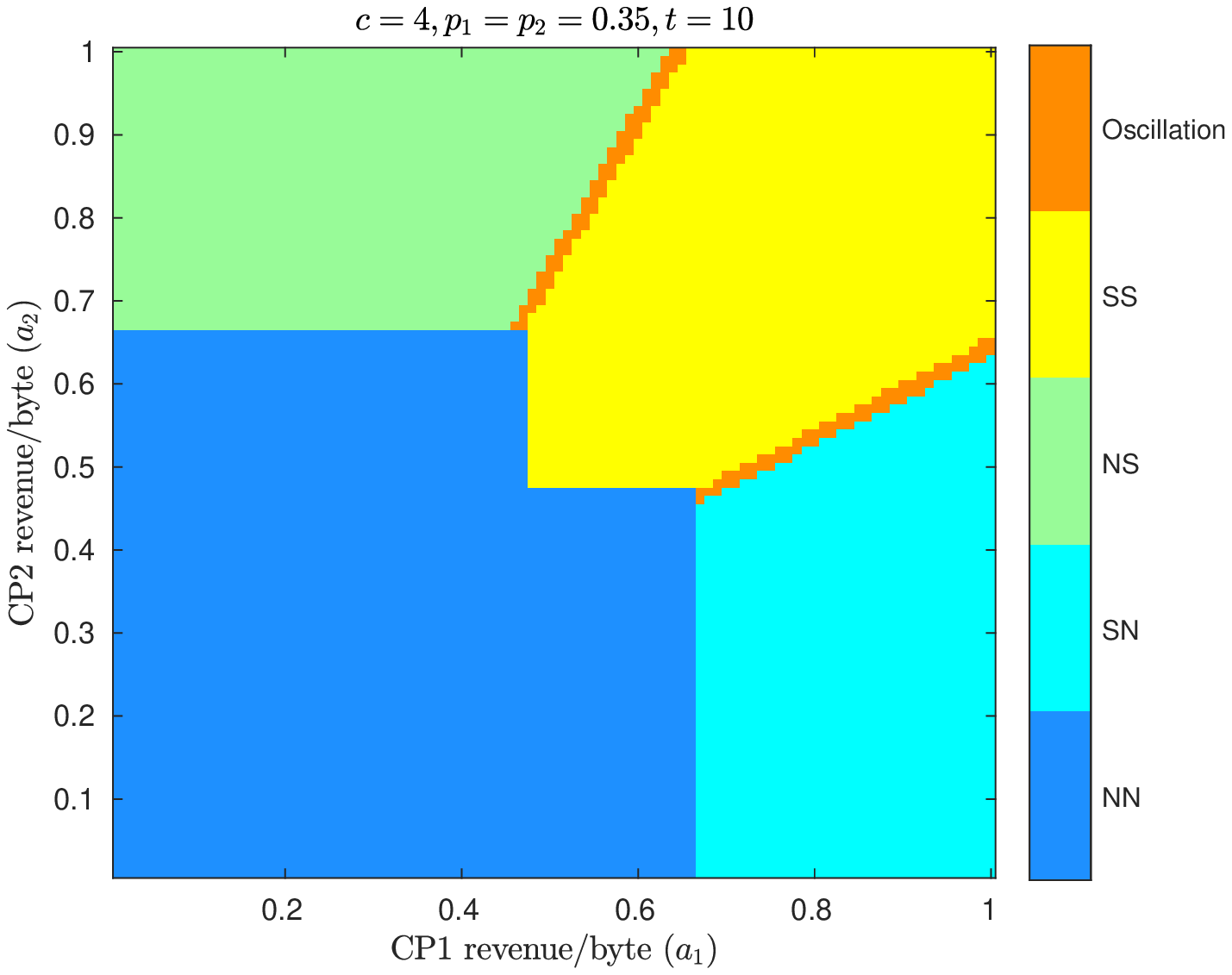}
      }
      \subfloat[]{
        \includegraphics{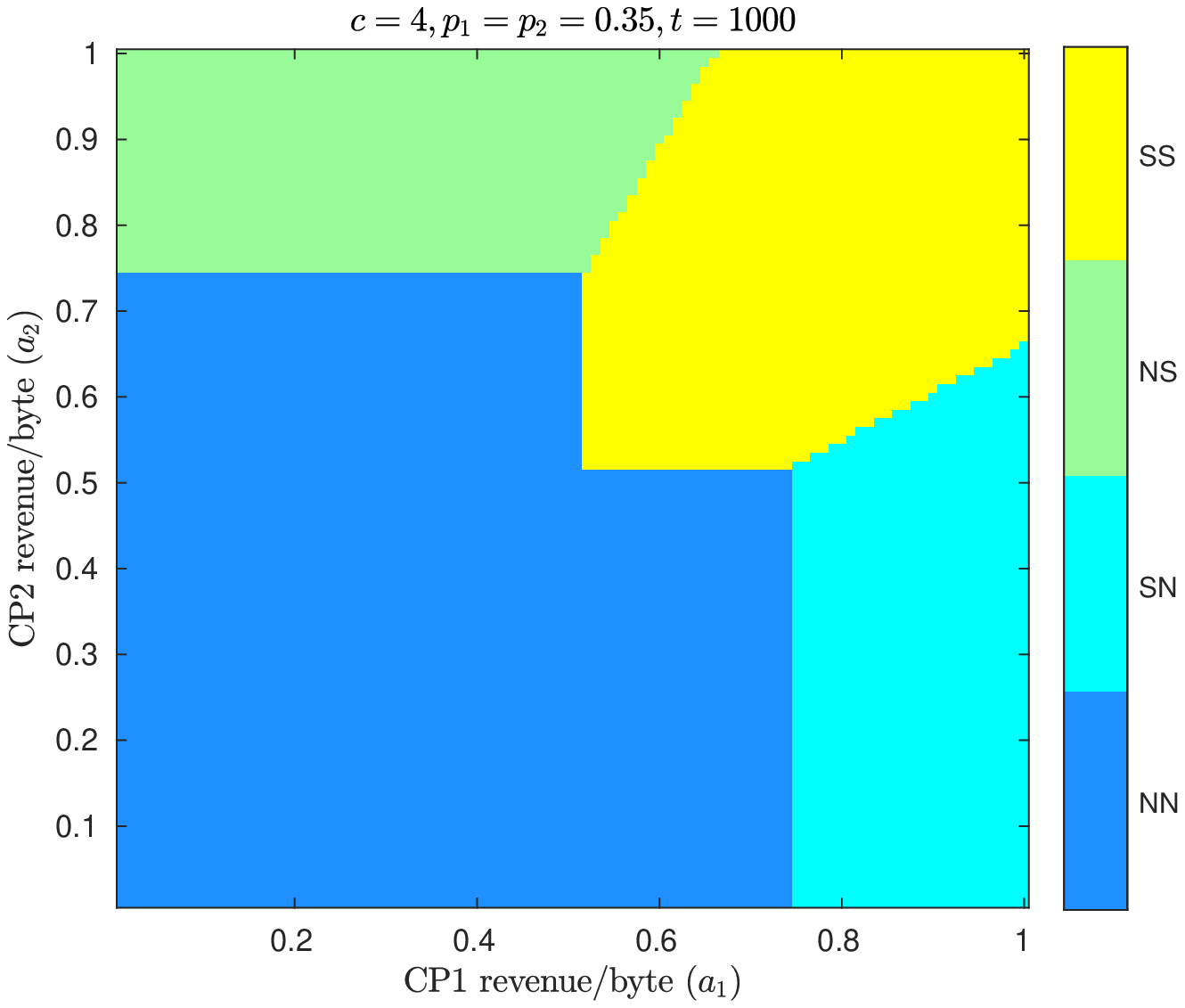}

       }
    }
    \caption{Limiting sponsorship configurations as a function of $a_1,a_2$ with $c=4$ and varying $t.$ (a) $t=3$ (b) $t=10$ (c) $t=1000$}
    \label{fig:market_c4_p35_t3_10_1000}
  \end{center}
\end{figure*}

We first report the (experimentally observed) limiting sponsorship
configurations from the best response dynamics over the
$a_1 \times a_2$ space. Interestingly, in all cases, we observe that
the equilibrium (when the dynamics converge) is symmetric across the
ISPs, i.e., both ISPs arrive at the same sponsorship
configuration. Moreover, these equilibrium configurations have the
same structural dependence on $a_1$ and $a_2$ as we saw in the
`single-step best response' characterization in Section~\ref{sec:opt_q1};
see Figure~\ref{fig:market_c4_p35_t3_10_1000}(a). When $a_1$ and $a_2$
are small, the equilibrium involves both ISPs in an NN configuration,
as expected. Moreover, when $a_1 \gg a_2$ or $a_2 \gg a_1,$ both ISPs
arrive at an equilibrium wherein the more profitable CP
sponsors. Finally, when $a_1$ and $a_2$ are comparable and large
enough, both ISPs induce both CPs to sponsor. We also observe that
there are certain intermediate regions in the $a_1 \times a_2$ space
where the best response dynamics oscillate. 

Next, we compare the limiting behavior of the best response dynamics
for different values of the transportation cost parameter $t;$ see
Figures~\ref{fig:market_c4_p35_t3_10_1000}(a)--\ref{fig:market_c4_p35_t3_10_1000}(c).
Recall that increasing $t$ implies increasing user stickiness, and
thus a diminishing dependence of one ISP's action on the other. Note
that as $t$ grows, the region of the $a_1 \times a_2$ where the ISPs
induce one or both CPs to sponsor shrinks. Interestingly, this is the
result of a \emph{prisoner's dilemma} between the ISPs: When $t$ is
small, i.e., when inter-ISP user churn is significant, each ISP has
the unilateral incentive to induce sponsorship even at small CP
revenue rates, to benefit from the resulting increase in its
subscriber base. However, once one ISP induces sponsorship, the other
ISP is also incentivised to induce sponsorship to recover its lost
market share. As a result, the ISPs arrive at an equilibrium that
leaves them both worse off; this will also be apparent from the plots
of ISP surplus reported later.

On the other hand, when $t$ is large, then each ISP's market share is
relatively insensitive to the other's actions, and so the ISPs induce
sponsorship only when it is mutually beneficial for them to do
so. This also explains why as $t$ becomes large, the region of the
$a_1 \times a_2$ space where the best response dynamics oscillate
diminishes.


\begin{figure*}
  \begin{center}
    \resizebox*{\linewidth}{!}{
      \subfloat[]{
	\includegraphics{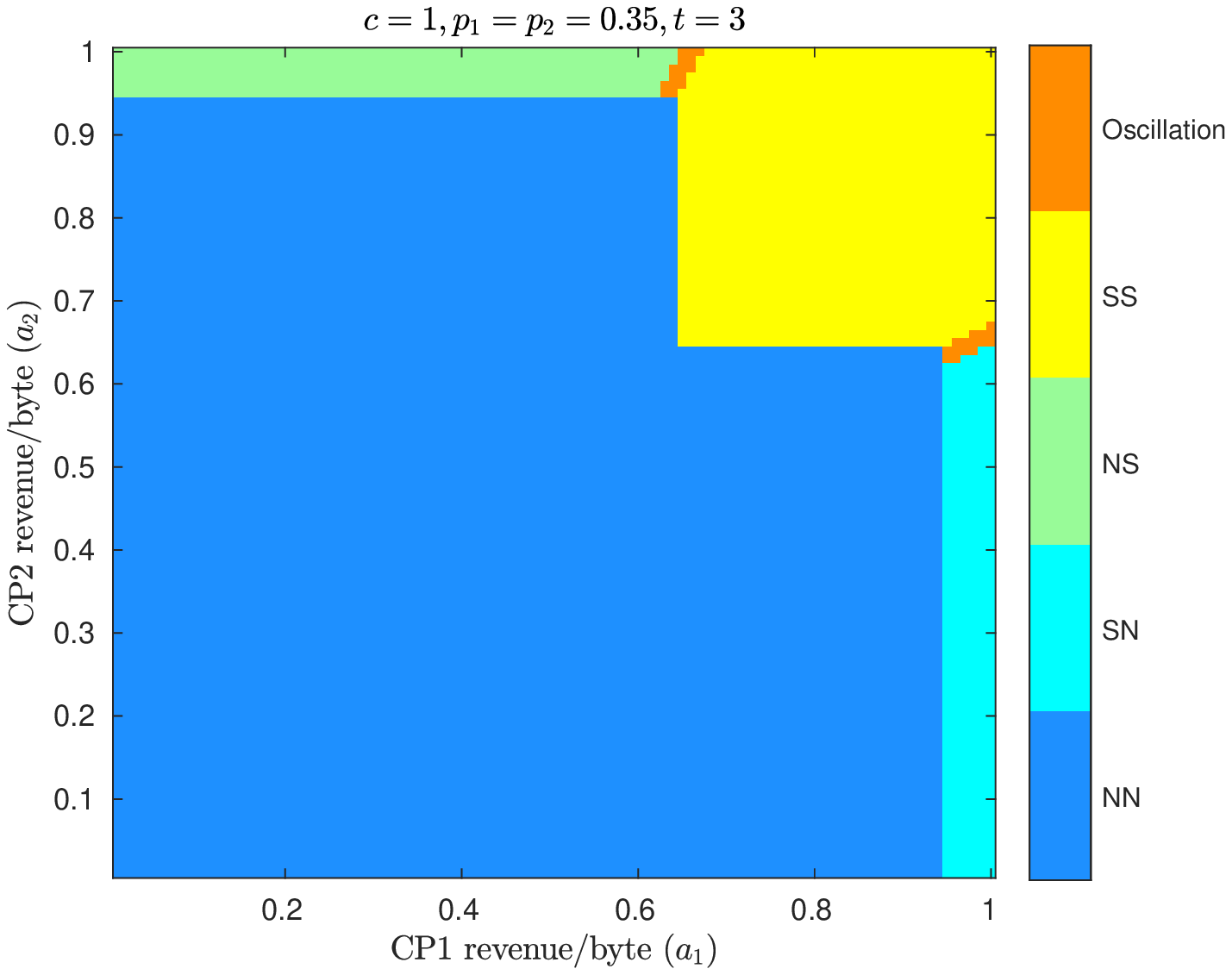}}
      \subfloat[]{
	\includegraphics{market_c4_p35_t3.eps}
      }
      \subfloat[]{
	\includegraphics{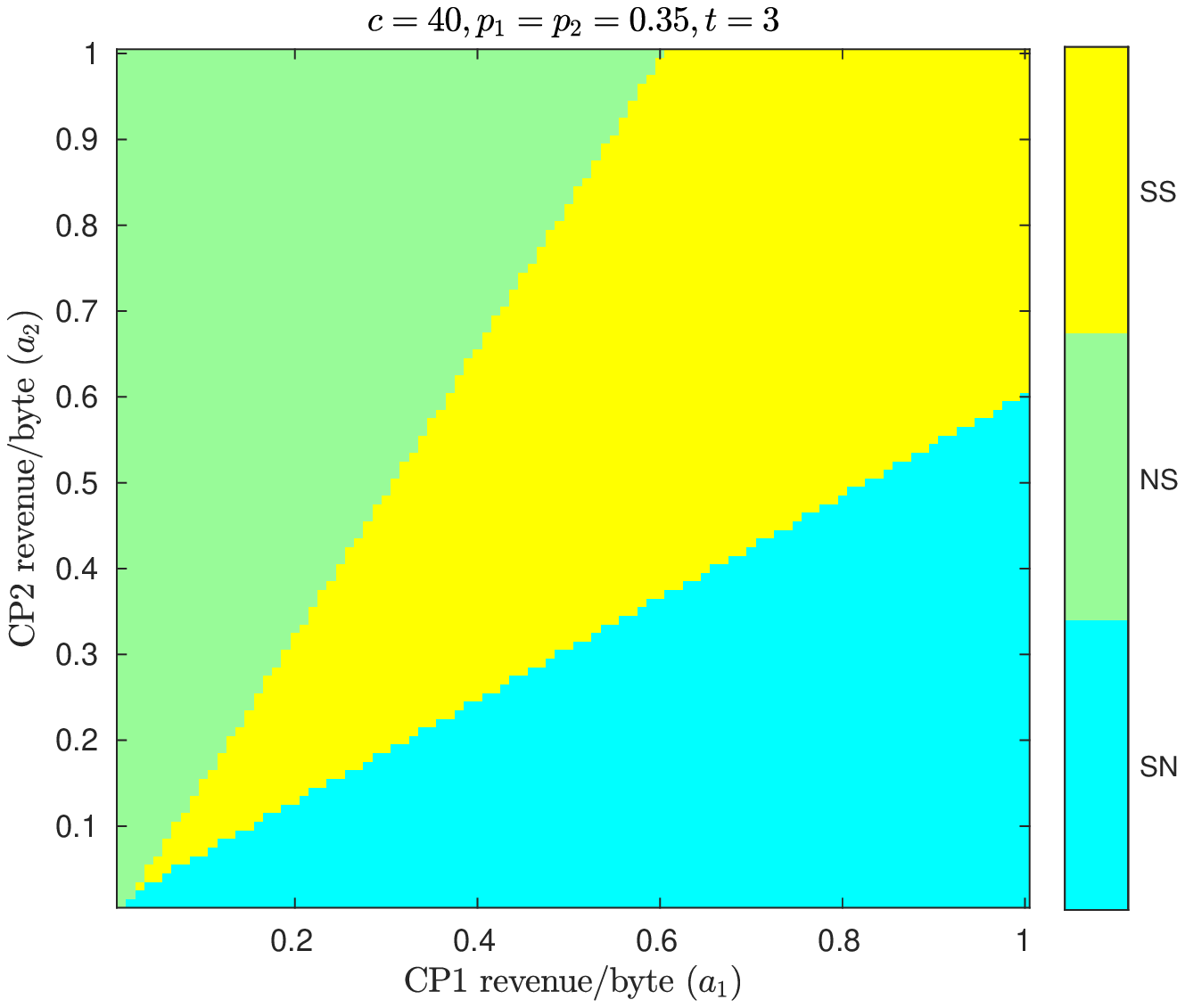}}
    }
    \caption{Limiting sponsorship configurations as a function of
      $a_1,a_2$ with $t=3$ and varying $c.$ (a) $c=1$ (b) $c=4$ (c)
      $c=40$}
    \label{fig:market_c1_40_p35_t3}
  \end{center}
\end{figure*}

Finally, we compare the limiting behavior of the best response
dynamics for different values of the `capacity to consume' $c;$ see
Figures~\ref{fig:market_c1_40_p35_t3}(a)--\ref{fig:market_c1_40_p35_t3}(c). Note
that when $c$ is small, there is only a modest growth in user-side
consumption from zero-rating. As a result, the equilibrium is NN on
both ISPs except when the CP revenue rates are really large. On the
other hand, when $c$ is large, ISPs induce sponsoring even at moderate
revenue rates to benefit from the increased data consumption from the
users.  

\subsection{Surplus}
\label{sec:symmetric_surplus}

Having explored the equilibrium sponsorship configurations that result
from alternating best response dynamics between ISPs, we now consider
the equilibrium surplus realized by the ISPs, the CPs, and the
users. Since a 3-d visualization of surplus in the $a_1 \times a_2$
space is hard to interpret, we use the parameterization
$(a_1,a_2)=(a,\rho a)$ for $\rho \in (0,1)$ (as in
Section~\ref{sec:opt_q1}). We first consider the case when $\rho$ is
small (i.e., CP2's revenue/byte is much less than that of CP1) and
then the case where $\rho$ is close to 1 (i.e., the revenue rates of
both CPs are comparable).

\begin{figure*}
  \begin{center}
    \resizebox*{\linewidth}{!}{
      \subfloat[]{
	\includegraphics{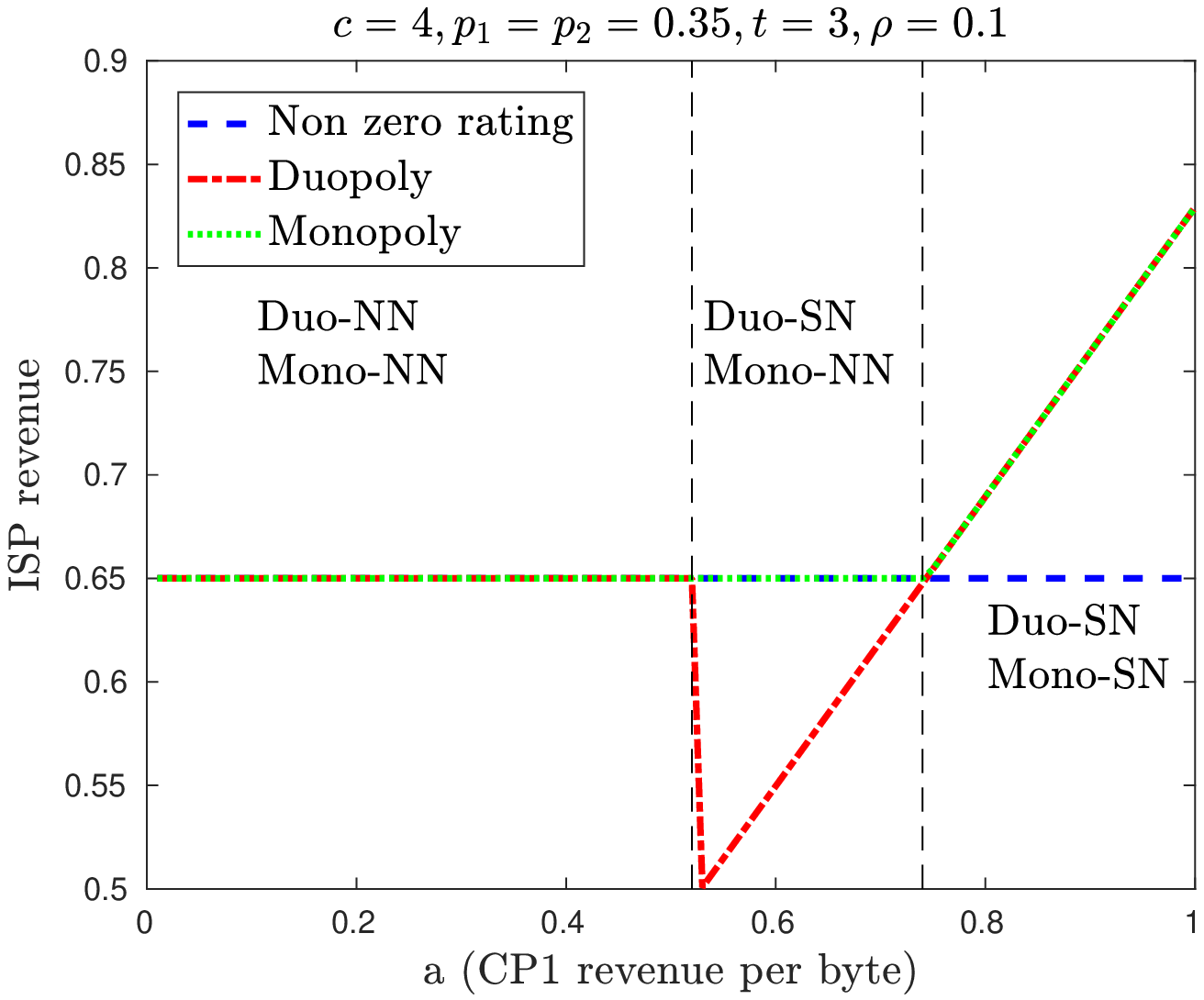}
      }
      \subfloat[]{
	\includegraphics{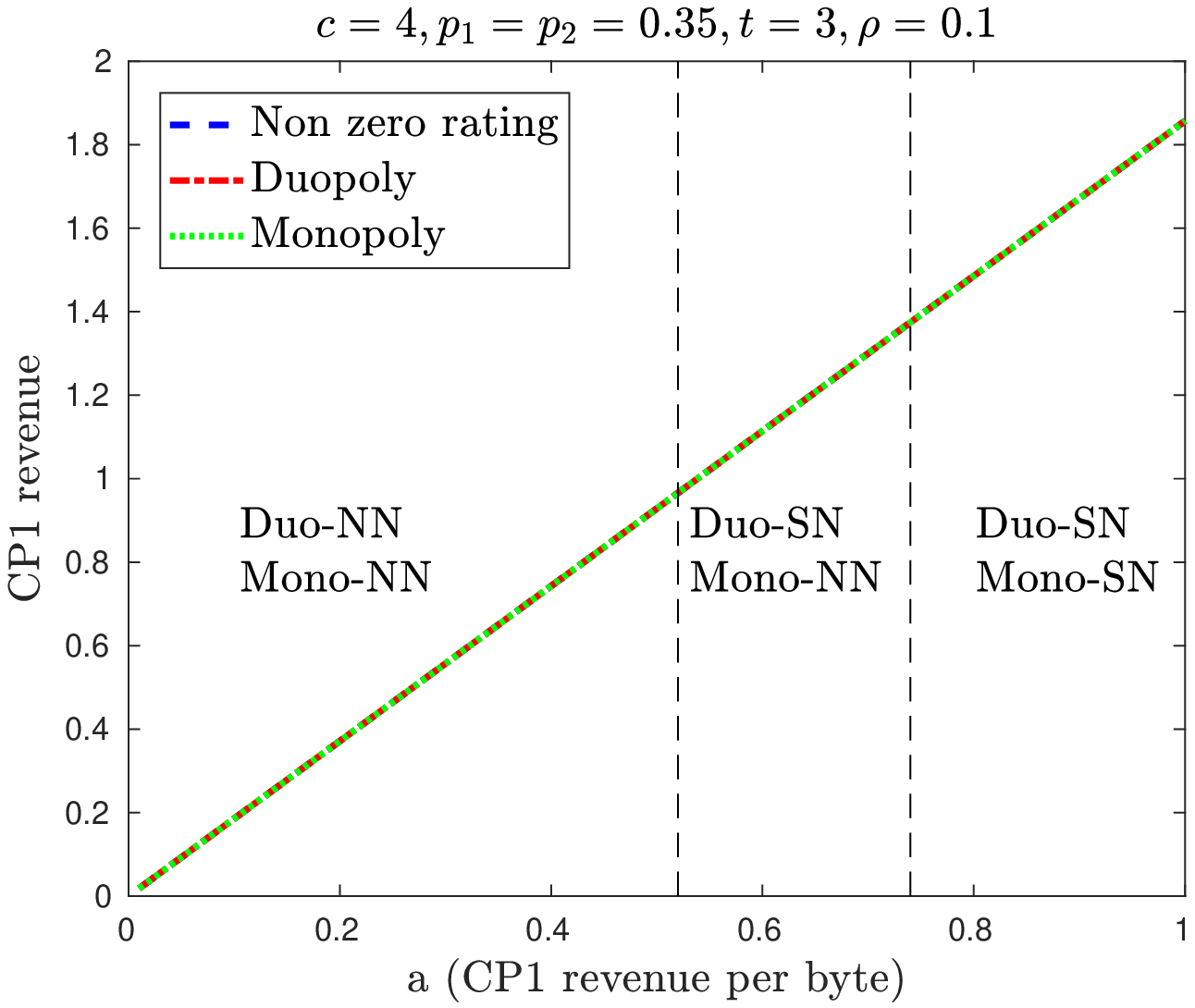}}
      \subfloat[]{
	\includegraphics{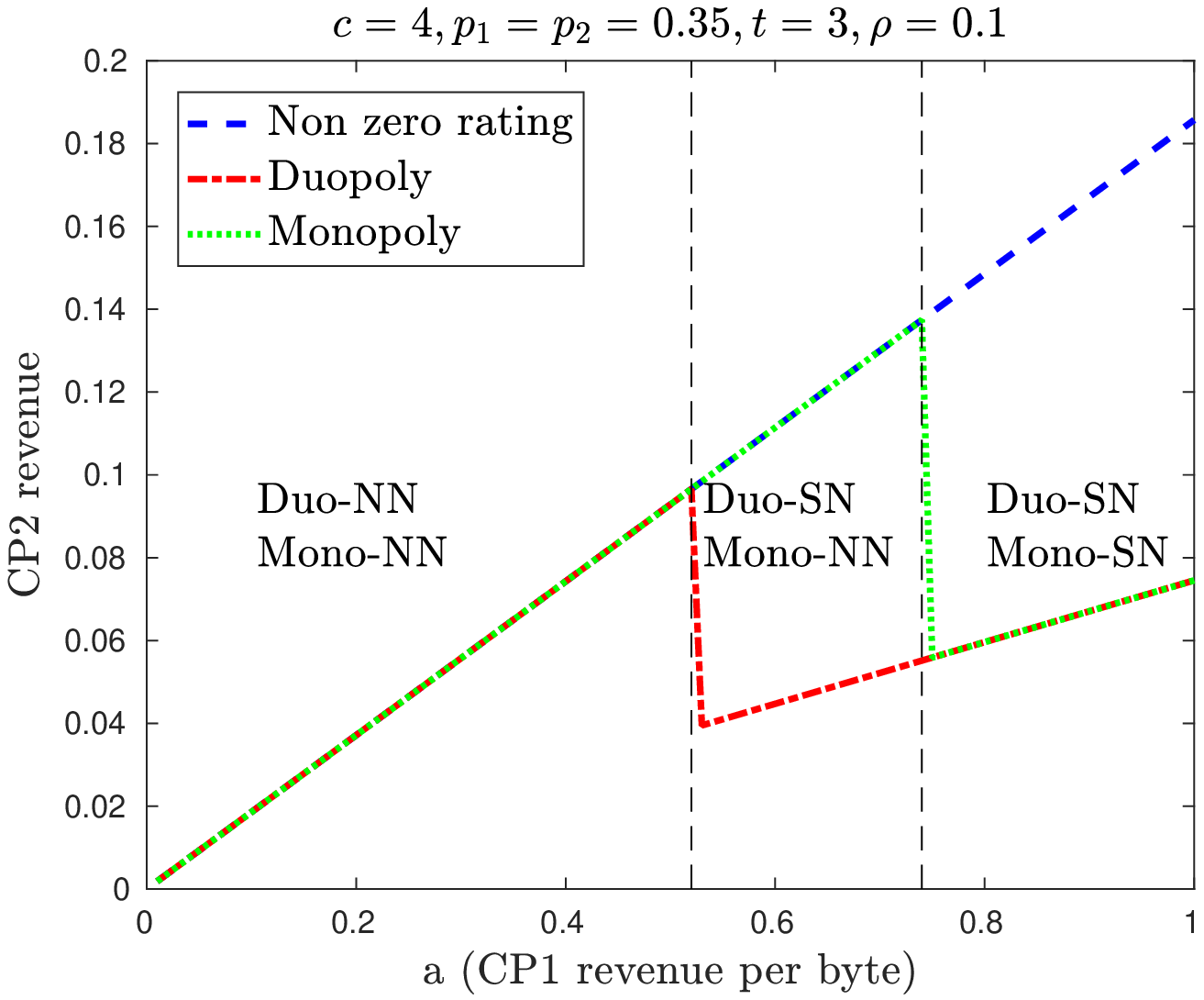}}
      \subfloat[]{
	\includegraphics{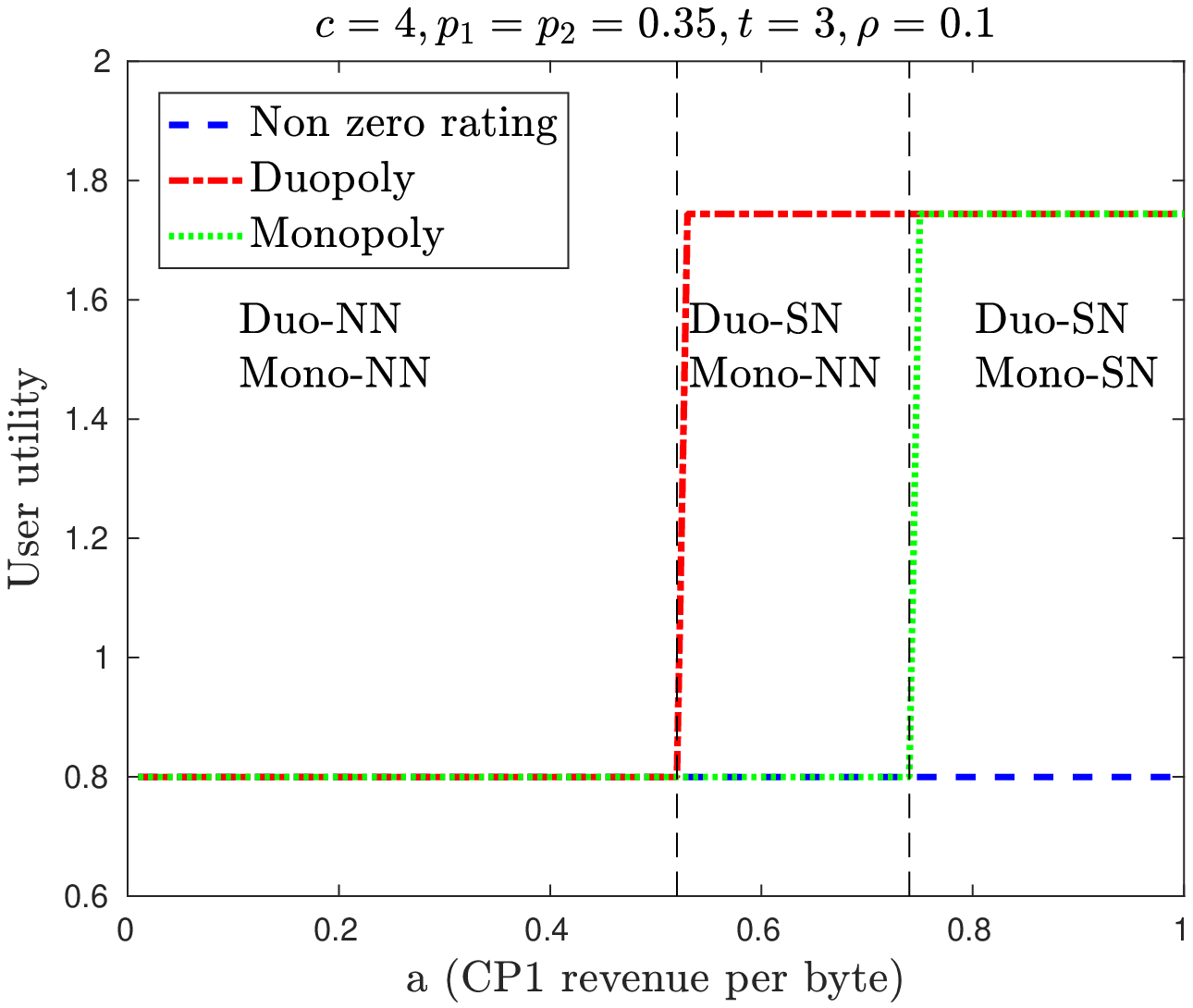}}
    }
    \caption{Surplus of various entities for $c=4,t=3,p =0.35$ and
      $\rho=0.1$ as a function of $a.$ (a) ISP revenue (b) CP1 revenue
      \\(c) CP2 revenue (d) User surplus}
    \label{fig:c4_p35_rho1}
  \end{center}
\end{figure*}

\noindent {\bf Small $\rho:$} Figure~\ref{fig:c4_p35_rho1} shows the
surplus of the ISPs (recall that since the equilibria we observe are
symmetric across the ISPs, both ISPs obtain the same surplus), CP1,
CP2, and the user base as a function of $a$ for $\rho=0.1$ and
$t = 3.$ From Figure~\ref{fig:market_c4_p35_t3_10_1000}(a), it is
clear that in this case, both ISPs induce an NN equilibrium for $a$
less than a certain threshold, and an SN equilibrium beyond this
threshold. We benchmark the equilibrium surplus under our model (ISP
\emph{duopoly}) with case where users are infinitely sticky (i.e.,
each ISP operates as a \emph{monopoly}) and the case where neither ISP
operates a zero-rating platform.

As was observed in Section~\ref{sec:symmetric_market}, competition
forces both ISPs to induce an SN configuration for smaller values of
$a$ as compared to the monopoly setting ($t \ra \infty$). This is
evident from the lower threshold (in $a$) for sponsorship as compared
to the monopoly setting. This \textit{prisoner's dilemma} between the
ISPs causes both ISPs to obtain a smaller profit compared with the
monopoly case for intermediate values of $a;$ see
Figure~\ref{fig:c4_p35_rho1}(a). For larger values of $a$ however, the
each ISP's surplus matches that in the monopoly case. The surplus of
the CP~1 (the sponsoring CP) remains the same under all three models,
in line with the conclusion of Theorem~\ref{thm:symmetric}; see
Figure~\ref{fig:c4_p35_rho1}(b). On the other hand, CP~2 (the
non-sponsoring CP) is worse off due to zero-rating, also in line with
Theorem~\ref{thm:symmetric}; see
Figure~\ref{fig:c4_p35_rho1}(c). Finally, we note that user surplus
gets enhanced due to zero-rating, as expected; see
Figure~\ref{fig:c4_p35_rho1}(d).

To summarize, we observe that except for intermediate values of $a,$
where competition forces both ISPs to induce sponsorship prematurely,
the surplus of all parties matches that in the monopoly case: the ISPs
are able to extract a considerable fraction of CP surplus, and neither
CP benefits from zero-rating. Indeed, as we prove below, for large
enough $a$ and small enough $\rho$ the monopoly configuration is
indeed a system equilibrium in our duopoly model.

\begin{theorem}
  Let $(a_1,a_2)= (a, \rho a)$ for $\rho \in (0,1).$ There exist
  thresholds $a_{SN} > 0$ and $\rho_{SN} > 0$ such that for
  $a > a_{SN}$ and $\rho < \rho_{SN},$ there exists $q(a)$ such that:
  \begin{enumerate}
  \item $(q(a),\text{SN},q(a),\text{SN})$ is a system
    equilibrium. Under this configuration, neither CP has the
    unilateral incentive to reverse its sponsorship decision on both
    ISPs. Moreover, CP~1 makes the same profit as it would in the
    absence of the zero-rating platforms, whereas CP~2 makes a profit
    less than or equal to that it would in the absence of the
    zero-rating platforms.
  \item In the monopoly setting ($t \ra \infty$), it is optimal for
    each ISP to induce an SN equilibrium by setting its sponsorship
    charge equal to $q(a).$
  \end{enumerate}
  \label{thm:isp_eq_a_large_rho_small}
\end{theorem}

Note that Theorem~\ref{thm:isp_eq_a_large_rho_small} does not prove
that for large enough $a$ and small enough $\rho,$ the best response
dynamics converge to the stated configuration. It merely establishes
that the configuration that the best response dynamics converge to in
our experiments is indeed a system equilibrium. In fact, it proves
that the observed configuration is an equilibrium in a stronger sense,
in that neither CP has the incentive to switch its sponsorship
decisions across both ISP platforms. The proof of
  Theorem~\ref{thm:isp_eq_a_large_rho_small} is presented in 
  Appendix~\ref{sec:a_large_rho_small_proof}.

\begin{figure*}
  \begin{center}
    \resizebox*{\linewidth}{!}{
      \subfloat[]{
	\includegraphics{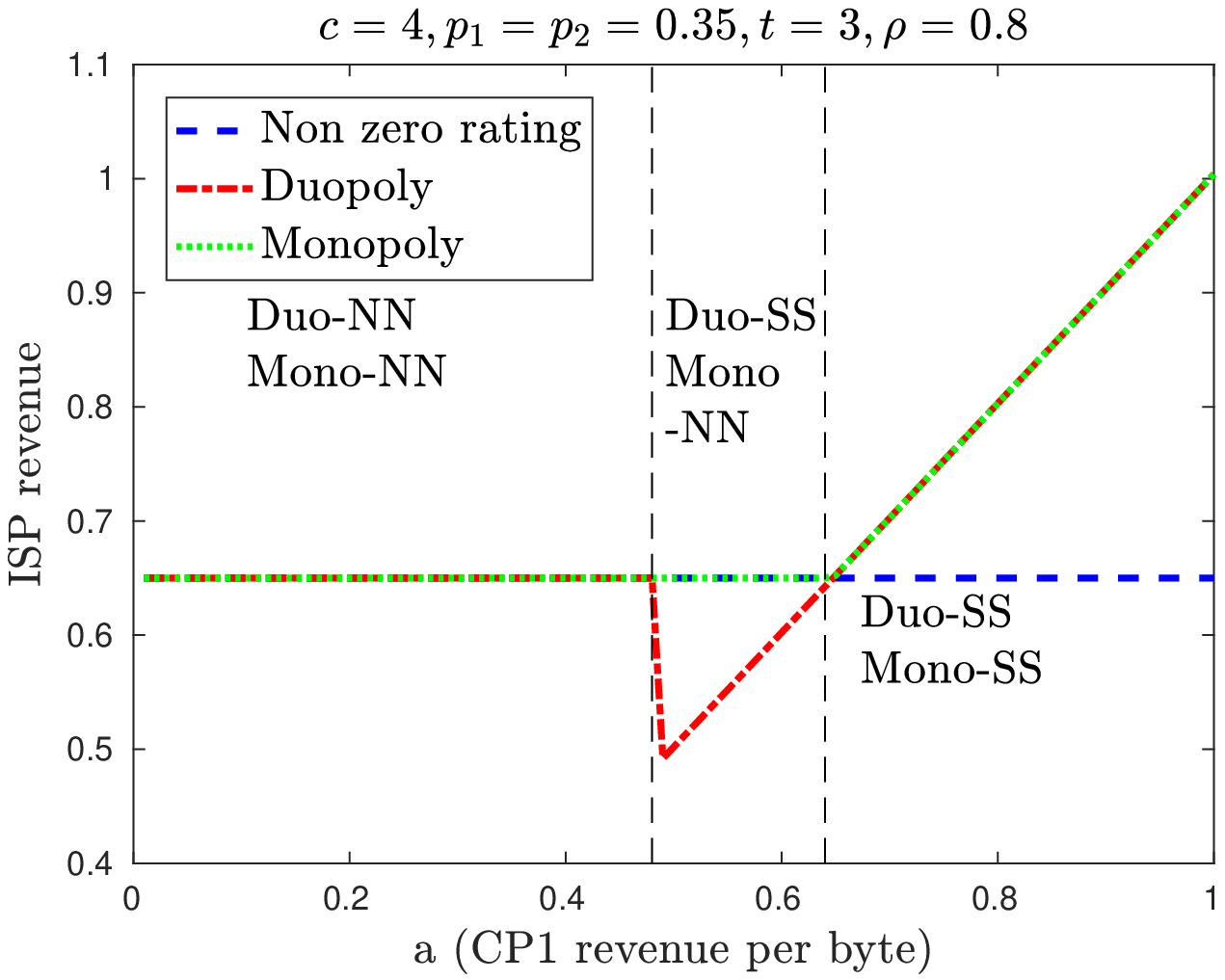}
      }
      \subfloat[]{
	\includegraphics{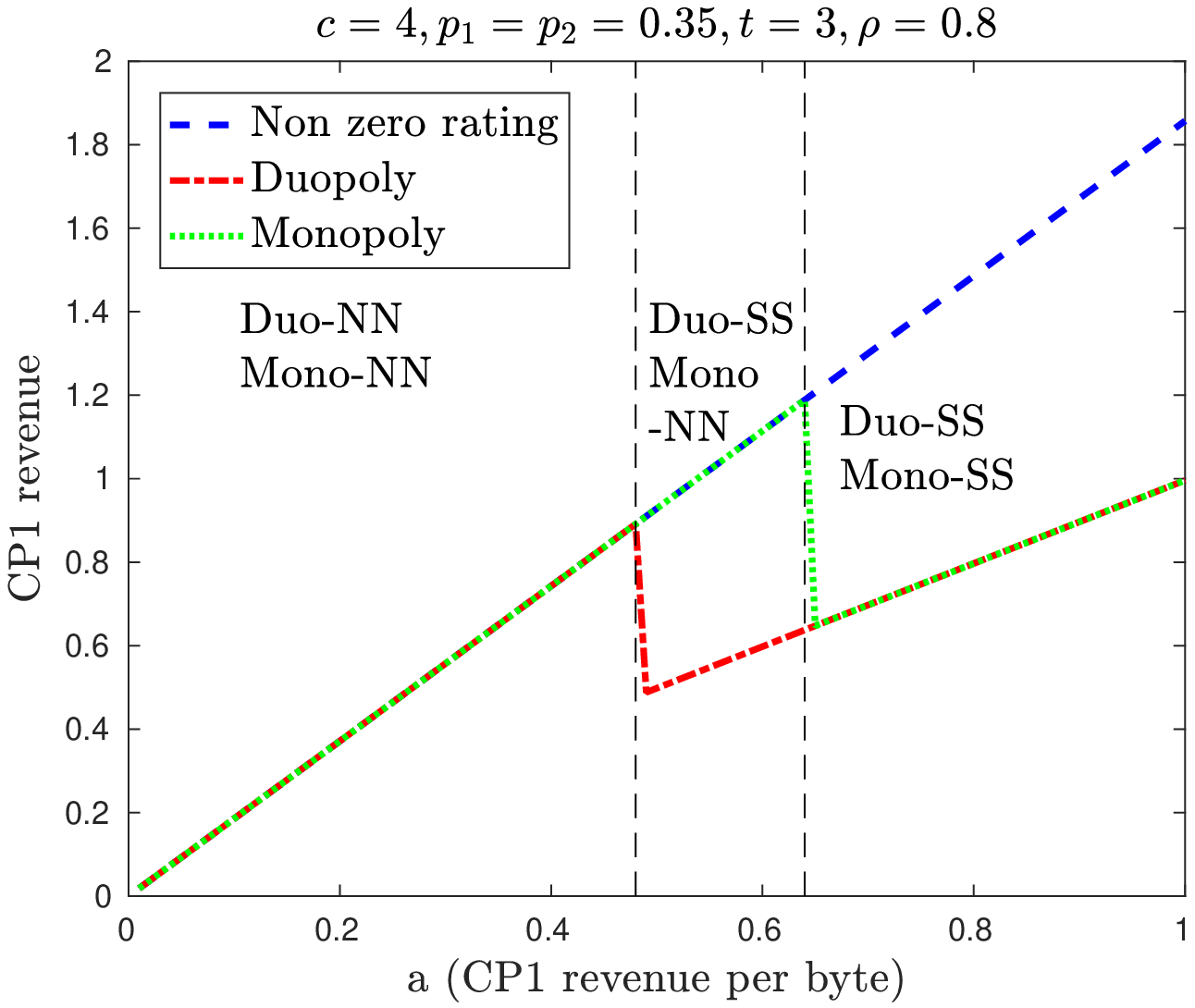}}
      \subfloat[]{
	\includegraphics{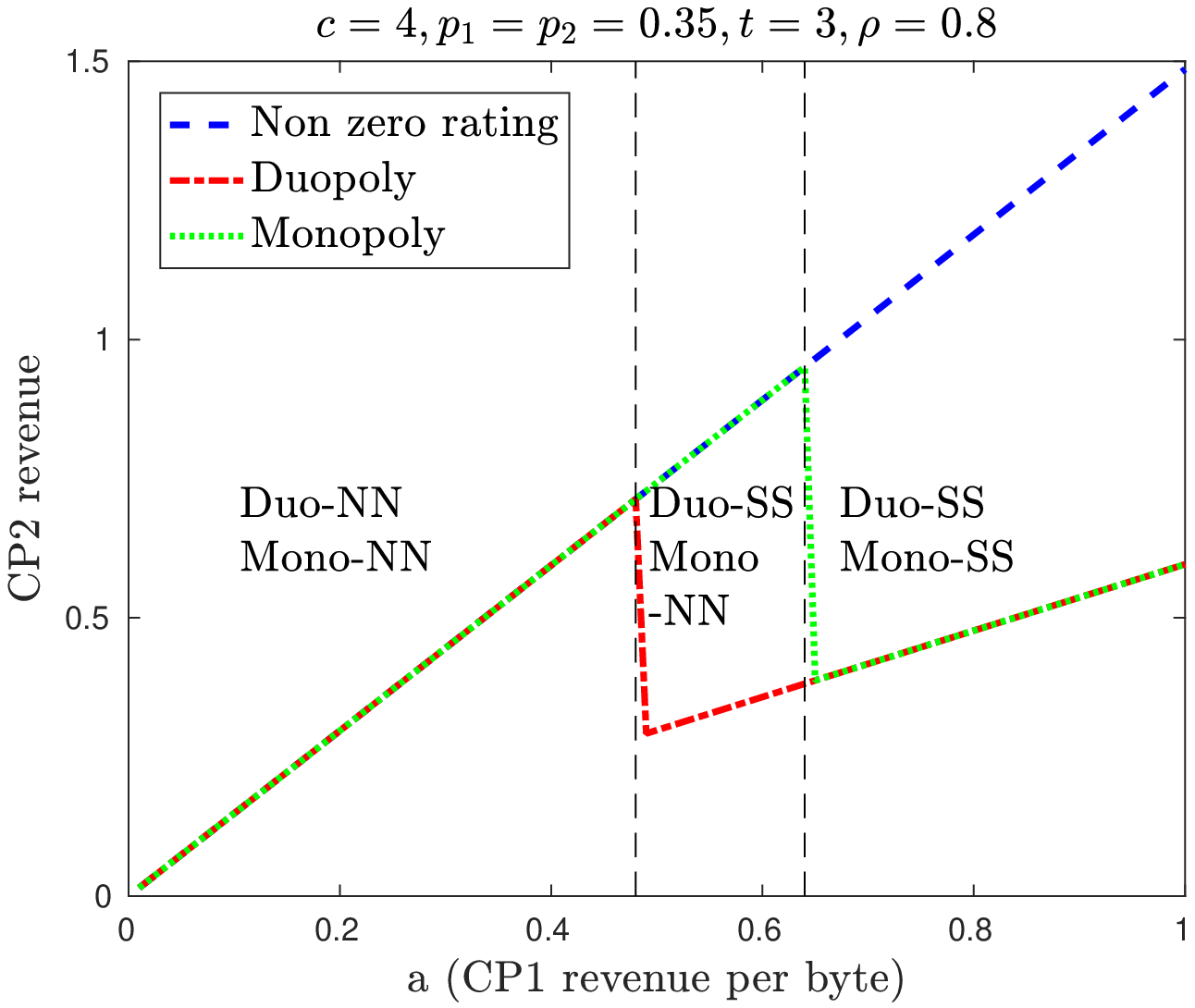}}
      \subfloat[]{
	\includegraphics{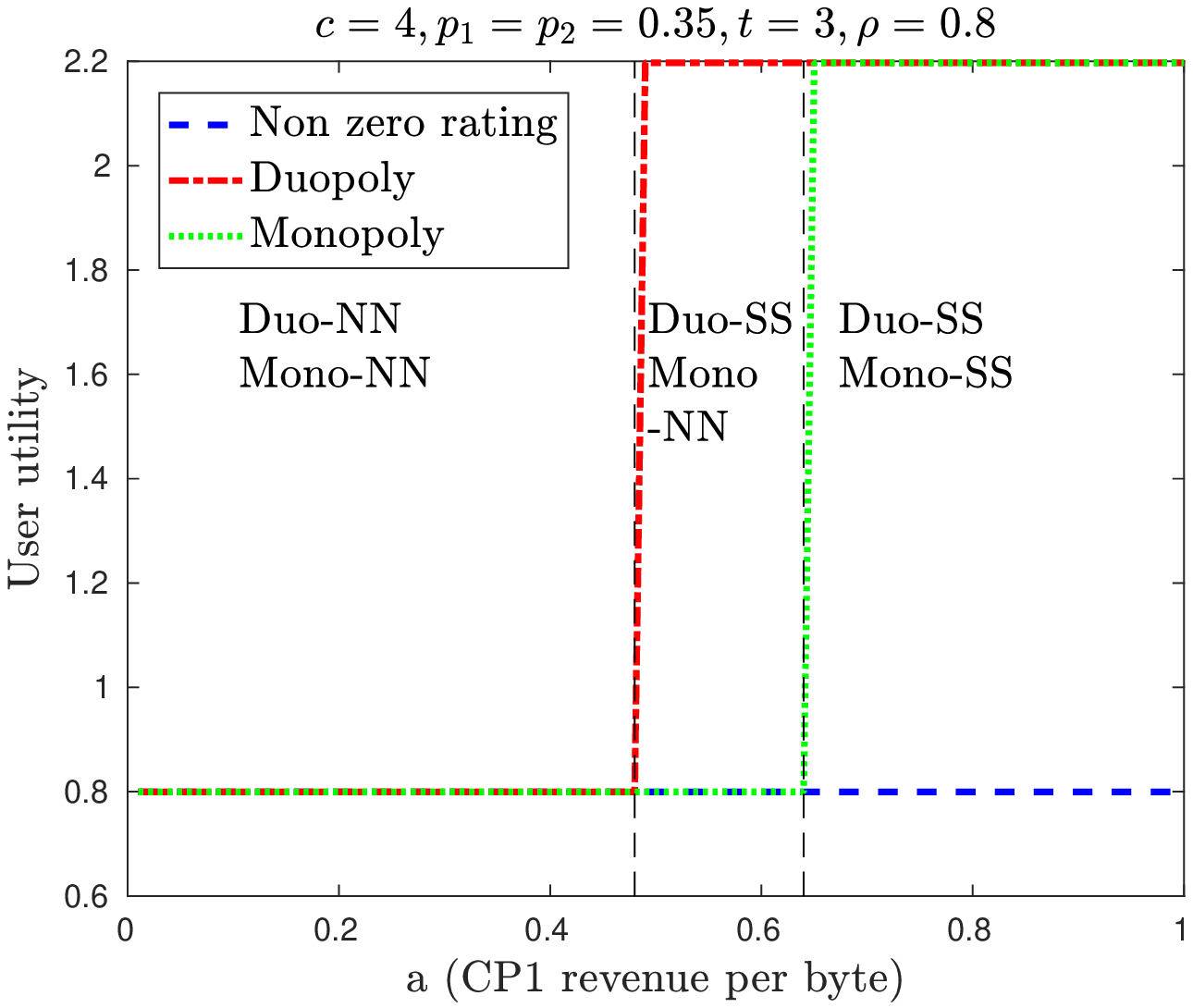}}
    }
    \caption{Surplus of various entities for $c=4,t=3,p=0.35$ and
      $\rho=0.8$ as a function of $a.$ (a) ISP revenue (b) CP1 revenue
      \\(c) CP2 revenue (d) User surplus}
    \label{fig:c4_p35_rho8}
  \end{center}
\end{figure*}

\noindent {\bf Large $\rho:$} Next, we consider the case where
$\rho = 0.8.$ Figure~\ref{fig:c4_p35_rho8} shows the surplus of
various entities as a function of $a.$ From
Figure~\ref{fig:market_c4_p35_t3_10_1000}(a), it is clear that in this
case, both ISPs induce an NN equilibrium for $a$ less than a certain
threshold, and an SS equilibrium beyond this threshold.
 
As before, we observe a prisoner's dilemma between the ISPs for
intermediate values of $a,$ where the ISP's enter into a mutually
sub-optimal sponsorship equilibrium; see
Figure~\ref{fig:c4_p35_rho8}(a). However, for larger values of $a,$
each ISP's surplus matches that in the monopoly
setting. Interestingly, this case also represents a prisoner's dilemma
between the CPs, wherein both CPs end up sponsoring for large enough
$a,$ and in the process end up worse off than if neither CP had
sponsored; see Figures~\ref{fig:c4_p35_rho8}(b)
and~\ref{fig:c4_p35_rho8}(c). Finally, we note as before that user
surplus is enhanced by sponsorship, more so than in the SN
configuration that emerges when $\rho$ is small; see
Figure~\ref{fig:c4_p35_rho8}(d).

As before, we prove that when $a$ and $\rho$ are large enough, the
observed equilibrium of the best response dynamics is indeed a system
equilibrium in our duopoly model. Moreover, under this configuration,
neither CP has the incentive to switch their sponsorship decision on
both ISPs.
	
\begin{theorem}
  Let $(a_1,a_2)= (a, \rho a)$ for $\rho \in (0,1).$ There exist
  thresholds $a_{SS}$ and $\rho_{SS}$ such that for $a > a_{SS}$ and
  $\rho > \rho_{SS},$ there exists $q(a,\rho)$ such that:
  \begin{enumerate}
  \item $q(a,\rho),\text{SS},q(a,\rho),\text{SS})$ is a system
    equilibrium. Under this configuration, neither CP has the
    unilateral incentive to reverse its sponsorship decision on both
    ISPs. Moreover, at least one CP makes a profit less than or equal
    to that it would in the absence of the zero-rating platforms.
  \item In the monopoly setting ($t \ra \infty$), it is optimal for
    each ISP to induce an SS equilibrium by setting its sponsorship
    charge equal to $q(a,\rho).$
  \end{enumerate}
    \label{thm:isp_eq_a_large_rho_large}
\end{theorem}
The proof of Theorem~\ref{thm:isp_eq_a_large_rho_large} is presented in 
	Appendix~\ref{sec:a_large_rho_large_proof}.

\noindent {\bf Intermediate $\rho:$} So far, we have seen that:
\begin{itemize}
\item For small enough $\rho$ and large enough $a,$ the limiting
  configuration under alternating best response dynamics is SN on both
  ISPs, which matches configuration under the monopoly setting.
\item For $\rho \approx 1$ and large enough $a,$ the limiting
  configuration under alternating best response dynamics is SS on both
  ISPs, which also matches configuration under the monopoly setting.
\end{itemize}

It is thus natural to ask what happens for intermediate values of
$\rho.$ In this section, we show that for intermediate values of
$\rho,$ a different type of \emph{prisoner's dilemma} can occur
between the ISPs, where both ISPs arrive at an SS configuration, even
though an SN configuration would be better for both ISPs.

\begin{figure*}
  \begin{center}
    \resizebox*{\linewidth}{!}{
      \subfloat[]{
	\includegraphics{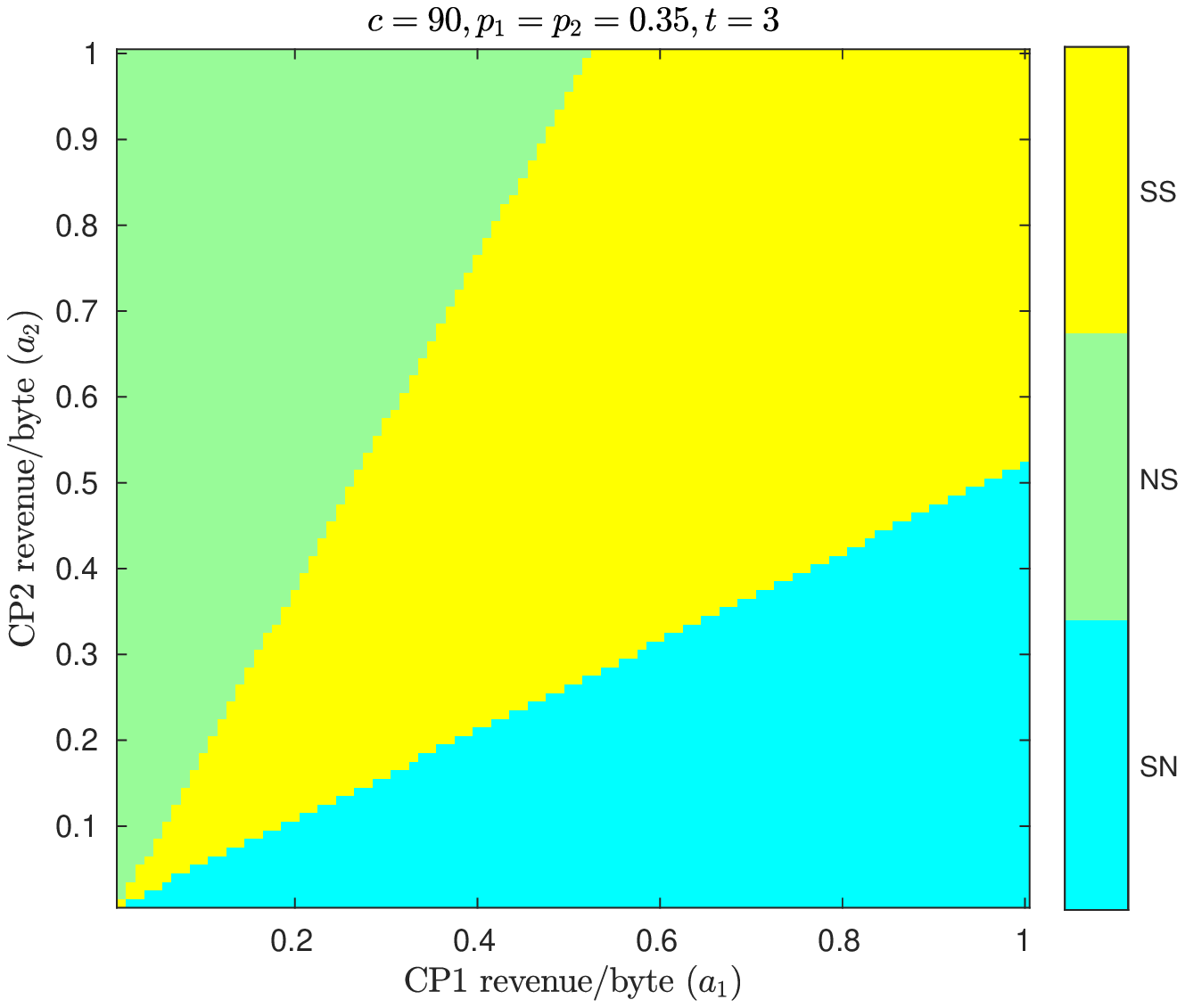}
      }
      \subfloat[]{
	\includegraphics{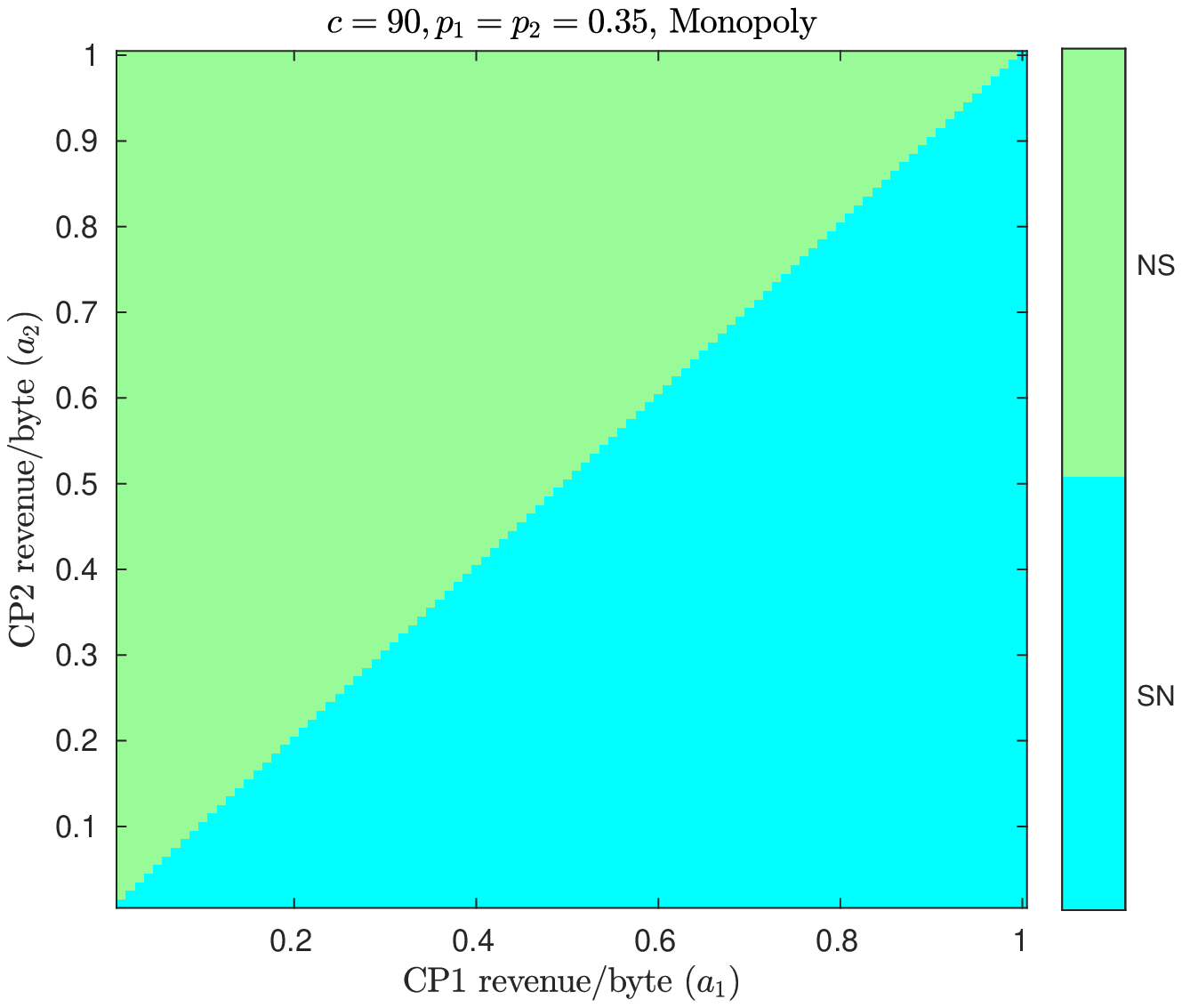}}
     \subfloat[]{
	\includegraphics{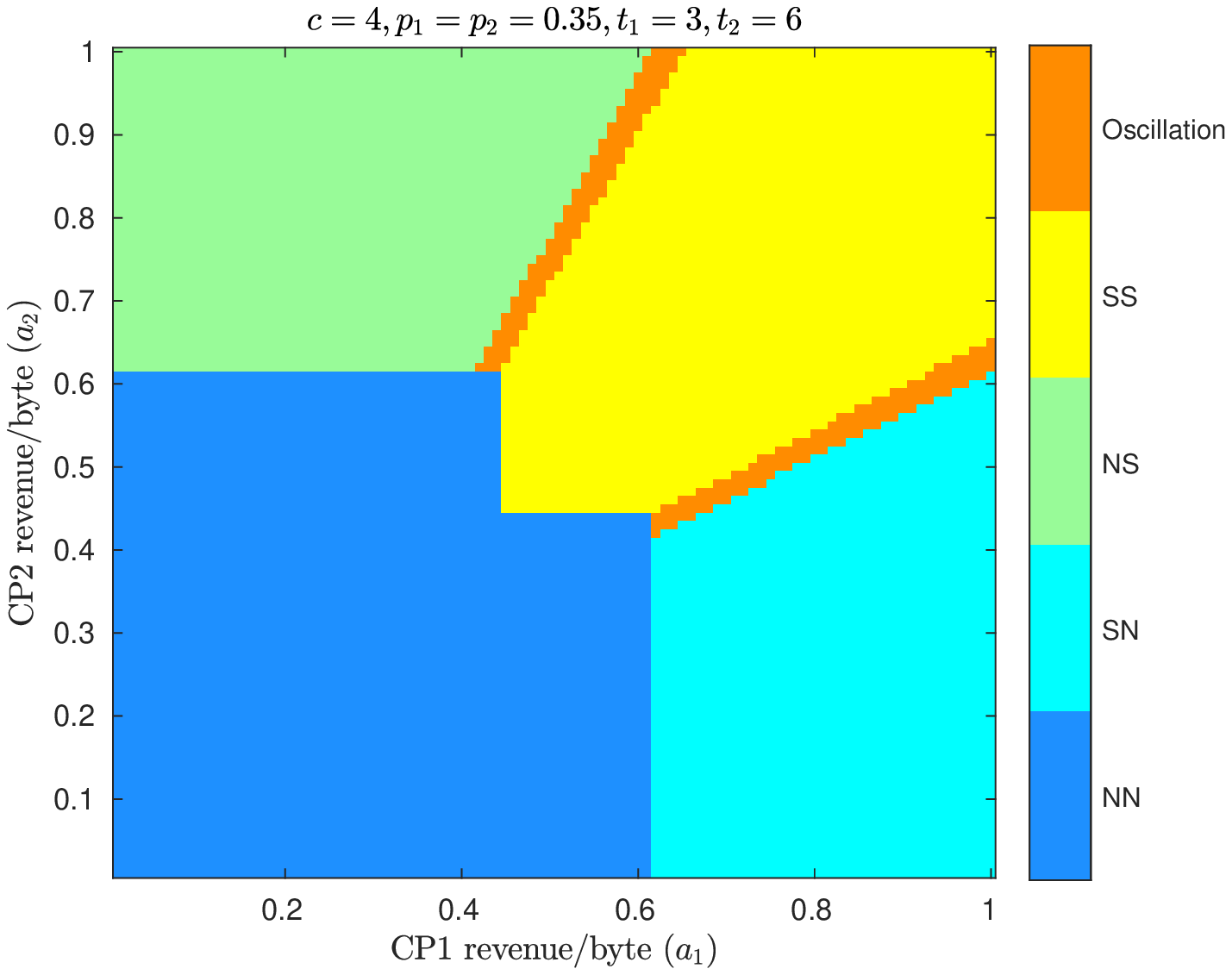}}
    }
    \caption{ISP configurations as a function of $a_1,a_2.$
      (a) Duopoly with $t=3$ for $c=90$ (b) Monopoly for $c=90$ (c) For asymmetric stickiness with $t_1=3, t_2=6,c=4.$ Initial ISP2 in NN.}
    \label{fig:market_c90_p35}
  \end{center}
\end{figure*}

To illustrate this most clearly, we set $c = 90.$
Figure~\ref{fig:market_c90_p35} shows the limiting ISP configurations
for duopoly and monopoly. We observe that for a range of $\rho,$ the
limiting duopoly configuration is SS on both ISPs, whereas in the
monopolistic setting, both ISPs prefer an SN equilibrium. This is a
different type of \emph{prisoner's dilemma} between the ISPs --- it is
optimal for both ISPs to operate an SN configuration. However, in this
state, each ISP has a unilateral incentive to switch to SS, in order to
gain a higher market share. However, once one ISP switches to SS, the
other ISP is also incentivised to switch to SS to regain its lost
market share, resulting in a mutually sub-optimal equilibrium.

\begin{figure*}
    \begin{center}
      \resizebox*{\linewidth}{!}{
	\subfloat[]{
	  \includegraphics{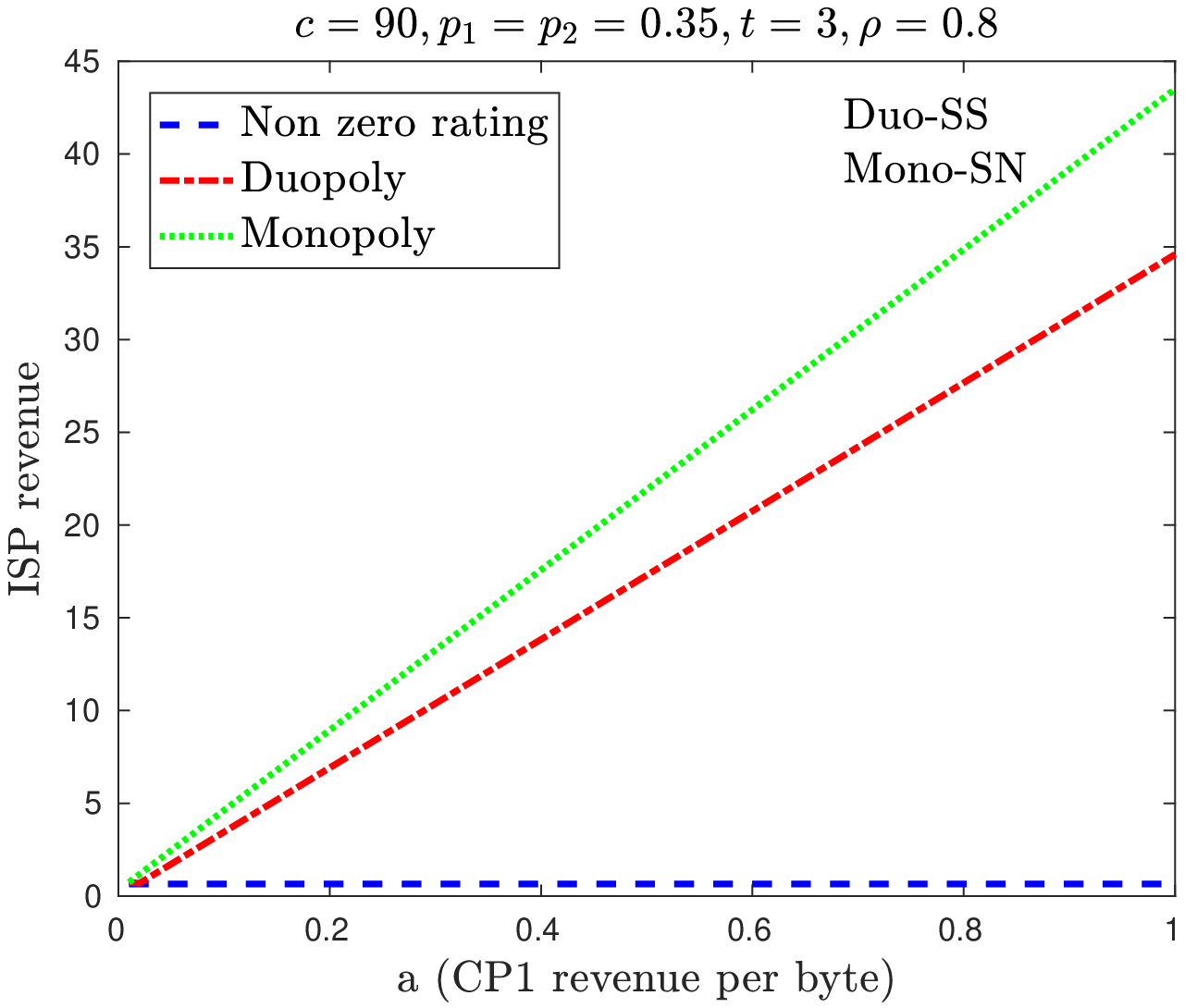}
	}
	\subfloat[]{
	  \includegraphics{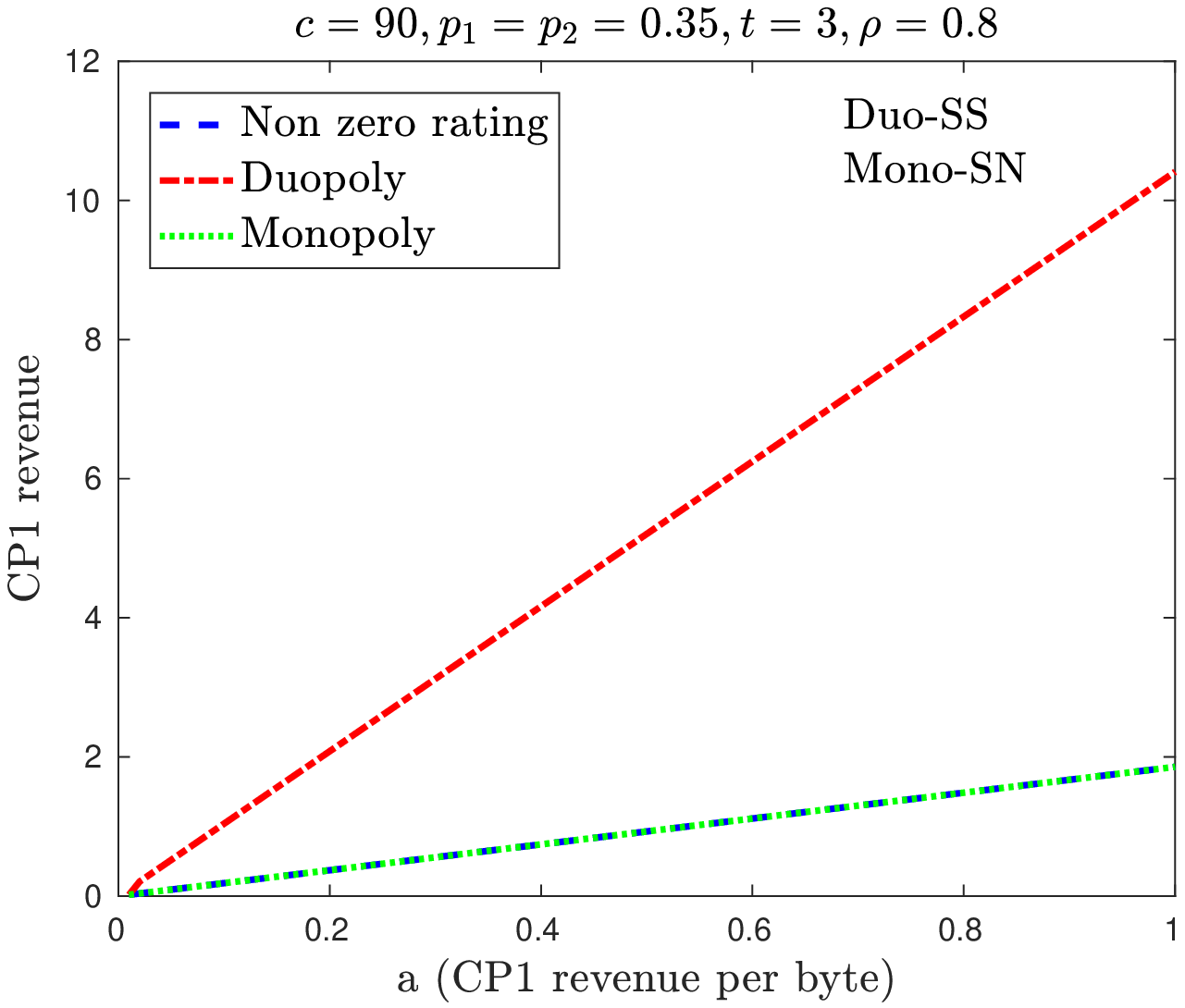}}
	\subfloat[]{
	  \includegraphics{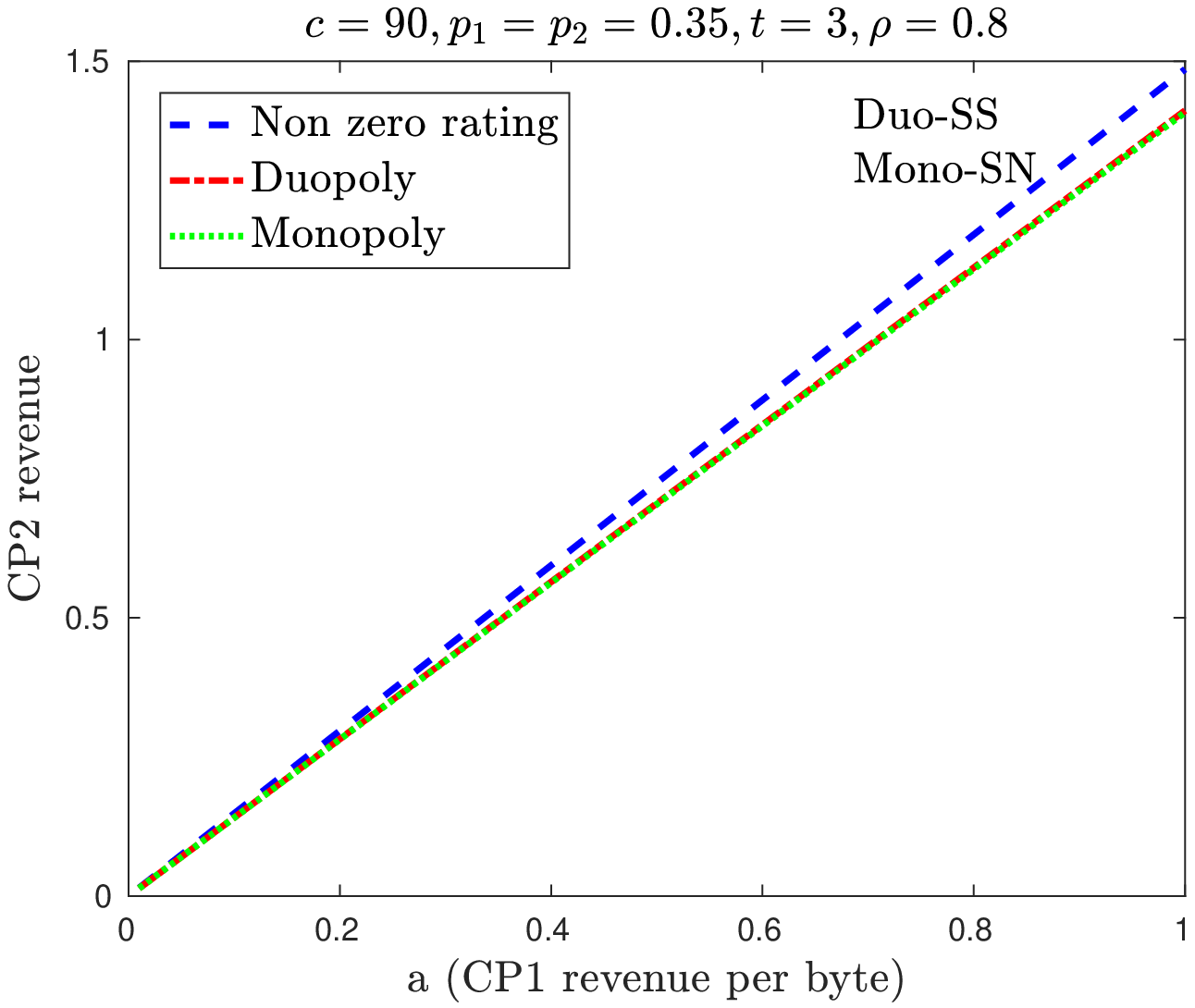}}
	\subfloat[]{
	  \includegraphics{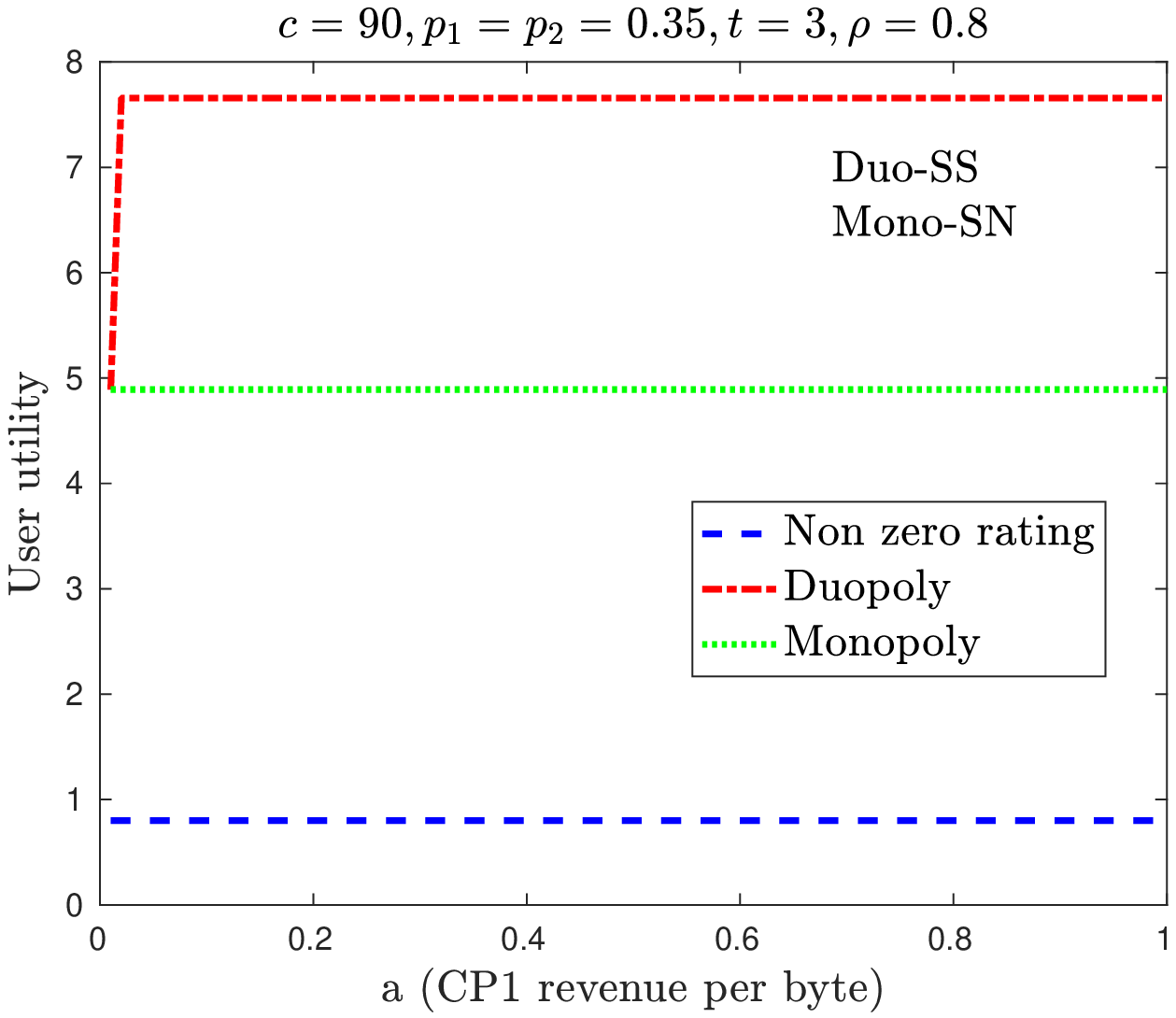}}
      }
      \caption{Surplus of various entities for $c=90,t=3,p_1=p_2=0.35$
        and $\rho=0.8$ as a function of $a$ (a) ISP revenue (b) CP1
        revenue (c) CP2 revenue (d) User surplus}
      \label{fig:c90_p35_rho8}
    \end{center}
  \end{figure*}

Figure~\ref{fig:c90_p35_rho8} shows surplus of various entities for
$\rho = 0.8.$ Note that ISP surplus is lower than that under the
monopoly setting. Indeed, CP~1 actually benefits from this
\emph{prisoner's dilemma} between the ISPs.

To summarize the key takeaways from this section, we see that
strategic interaction between ISPs, as captured by alternating best
response dynamics, can result in:
\begin{itemize}
 
\item Identical configuration to the monopoly setting. In this case,
  the ISPs are not affected by inter-ISP competition. Both ISPs manage
  to extract a portion of CP-side surplus, and at least one CP is
  worse off due to zero-rating.

 \item Prisoner's dilemma between the ISPs. This occurs for (i)
   intermediate values of $a,$ with small $\rho$ or $\rho \approx 1,$
   and (ii) intermediate values of $\rho.$ In this case, the ISPs are
   hurt by inter-ISP competition. However, the CPs do not necessarily
   benefit even in this case; at least one CP still ends up worse off
   due to zero-rating.
\end{itemize}

\section{Asymmetric stickiness}
\label{sec:asymmetric}
In this section, we consider a generalization of the Hotelling model,
where user-stickiness is asymmetric across the two ISPs. This might
capture, for example, the scenario where one ISP enjoys higher
customer loyalty than the other. We show that while the equilibria of
best response dynamics between the ISPs are qualitatively similar to
those in Section~\ref{sec:examples}, the ISP that enjoys a higher user
stickiness benefits more than the other ISP.

The generalized Hotelling model is parameterized by two parameters
$t_1$ and $t_2.$ Under this model, for sponsorship configuration $M_j$
on ISP~$j,$ the fraction of subscribers of ISP~1, denoted
$x_{M_1}^{M_2},$ is given by
$$u^{M_1} - t_1 x_{M_1}^{M_2} = u^{M_2} - t_2 (1-x_{M_1}^{M_2}).$$
To ensure a meaningful solution, we assume
$t_1,t_2 > u^{SS} - u^{NN}.$ Note that if $t_1 < t_2,$ users incur a
lower transportation cost to ISP~1 as compared to ISP~2, implying that
ISP~1 enjoys a higher user stickiness than ISP~2.

\begin{figure*}
	\begin{center}
		\resizebox*{\linewidth}{!}{
			\subfloat[]{
				\includegraphics{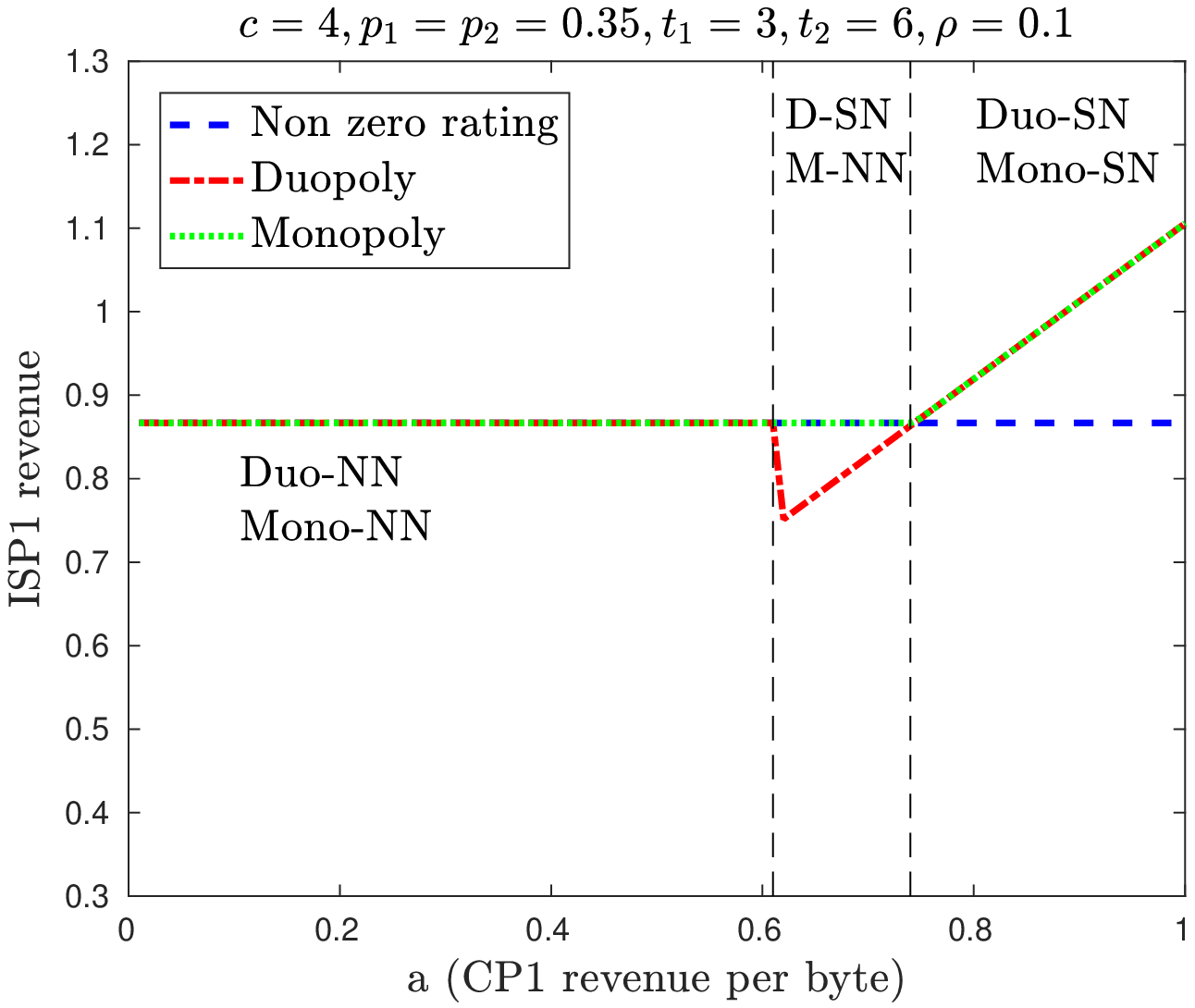}
			}
		    \subfloat[]{
		    	\includegraphics{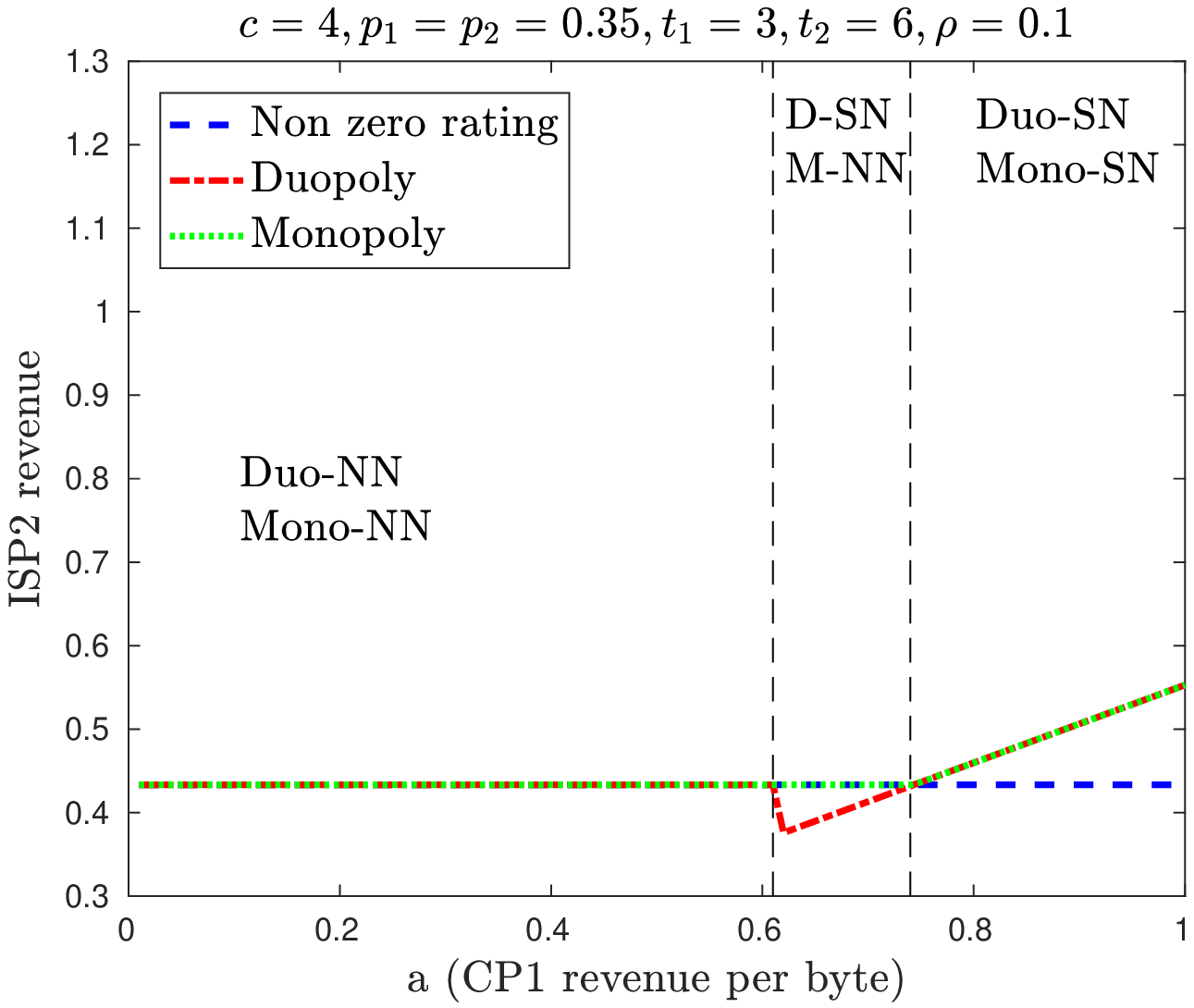}}
			\subfloat[]{
				\includegraphics{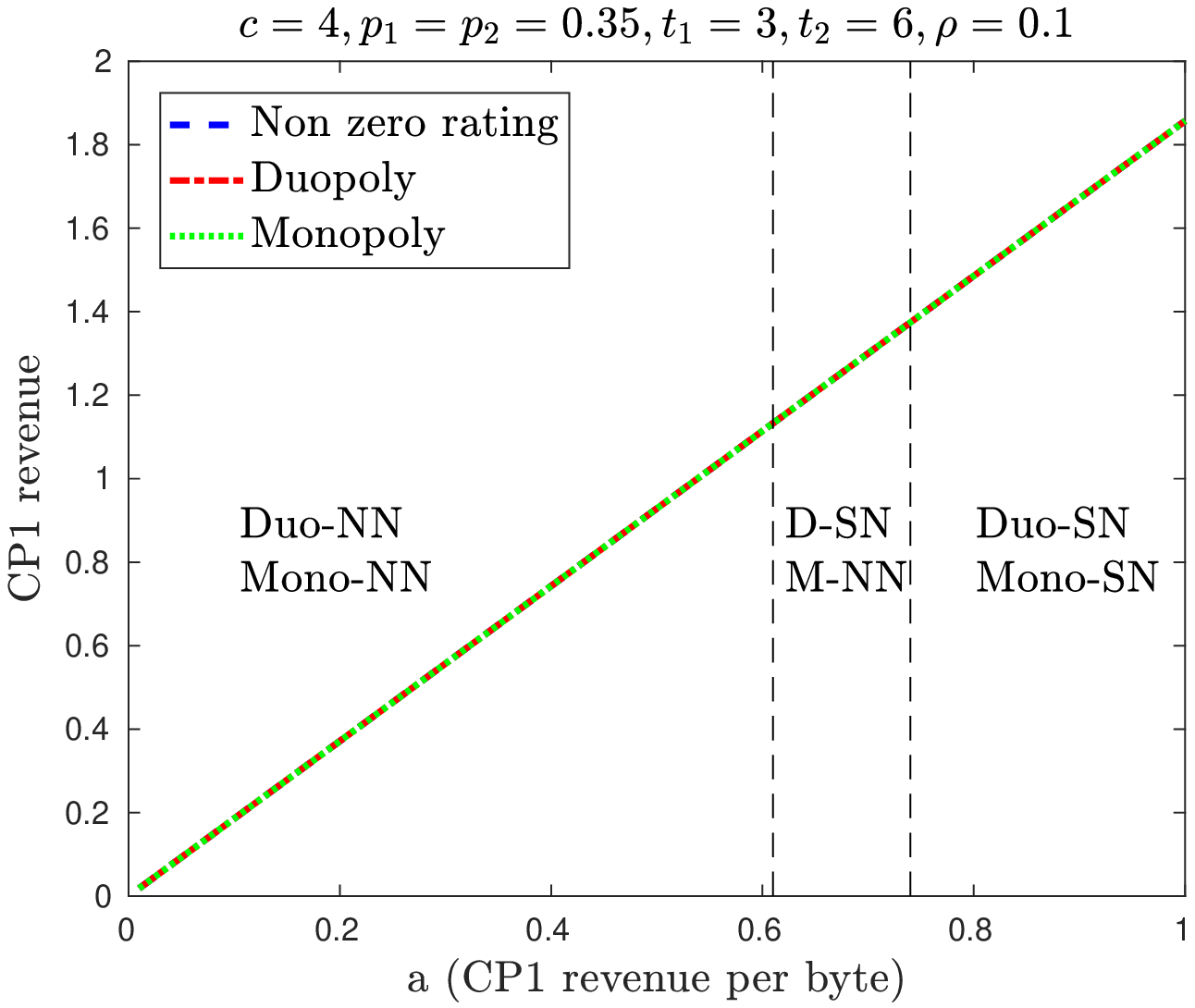}}
			\subfloat[]{
				\includegraphics{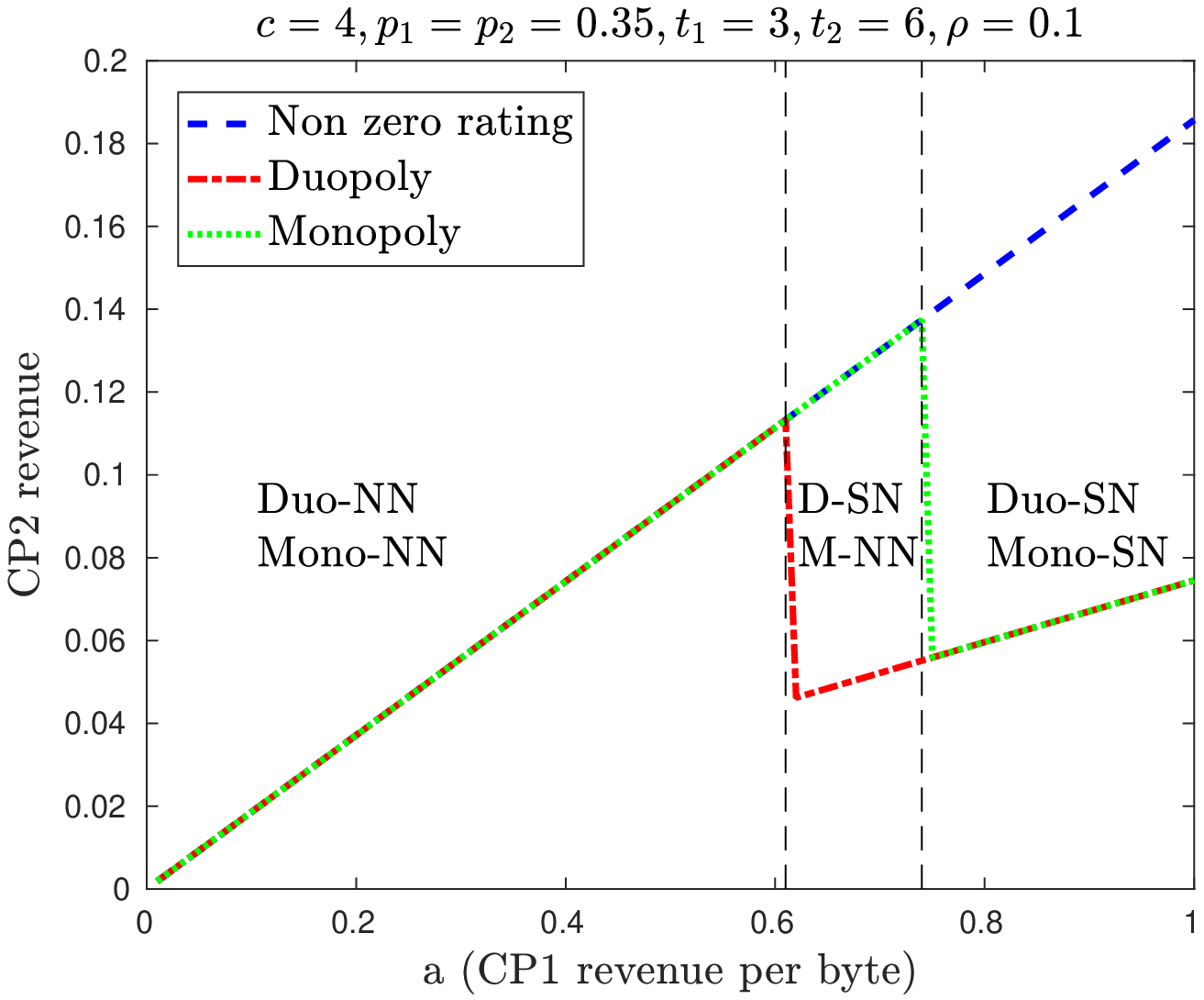}}
		}
		\caption{Surplus of various entities for $c=4,t_1=3,t_2=6,p_1=p_2=0.35$ and $\rho=0.1$ as a function of $a$ (a) ISP1 revenue (b) ISP2 revenue (c) CP1 revenue (d) CP2 revenue}
		\label{fig:assym_c4_p35_rho1}
	\end{center}
\end{figure*}

\begin{figure*}
	\begin{center}
		\resizebox*{\linewidth}{!}{
			\subfloat[]{
				\includegraphics{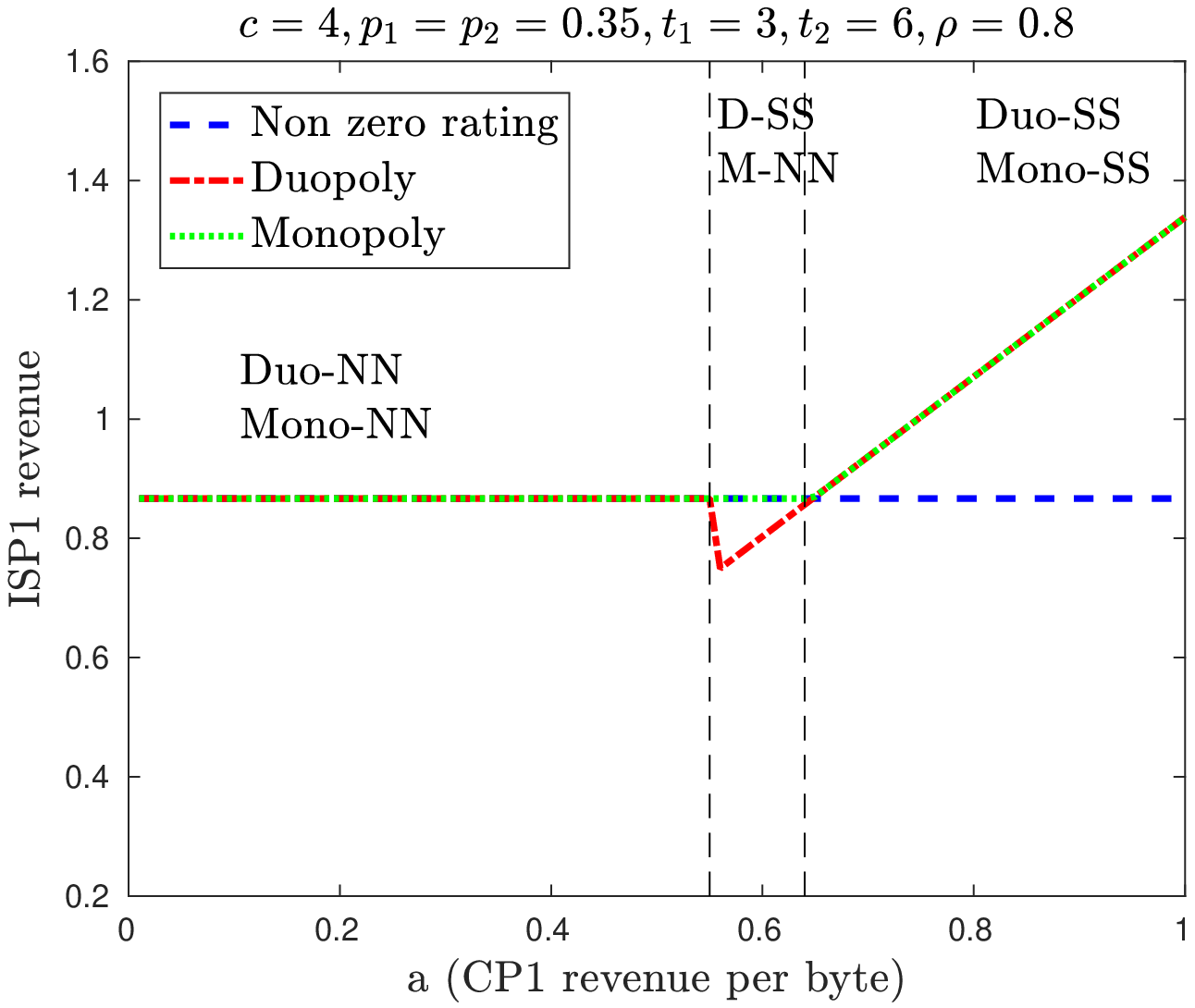}
			}
		    \subfloat[]{
		    	\includegraphics{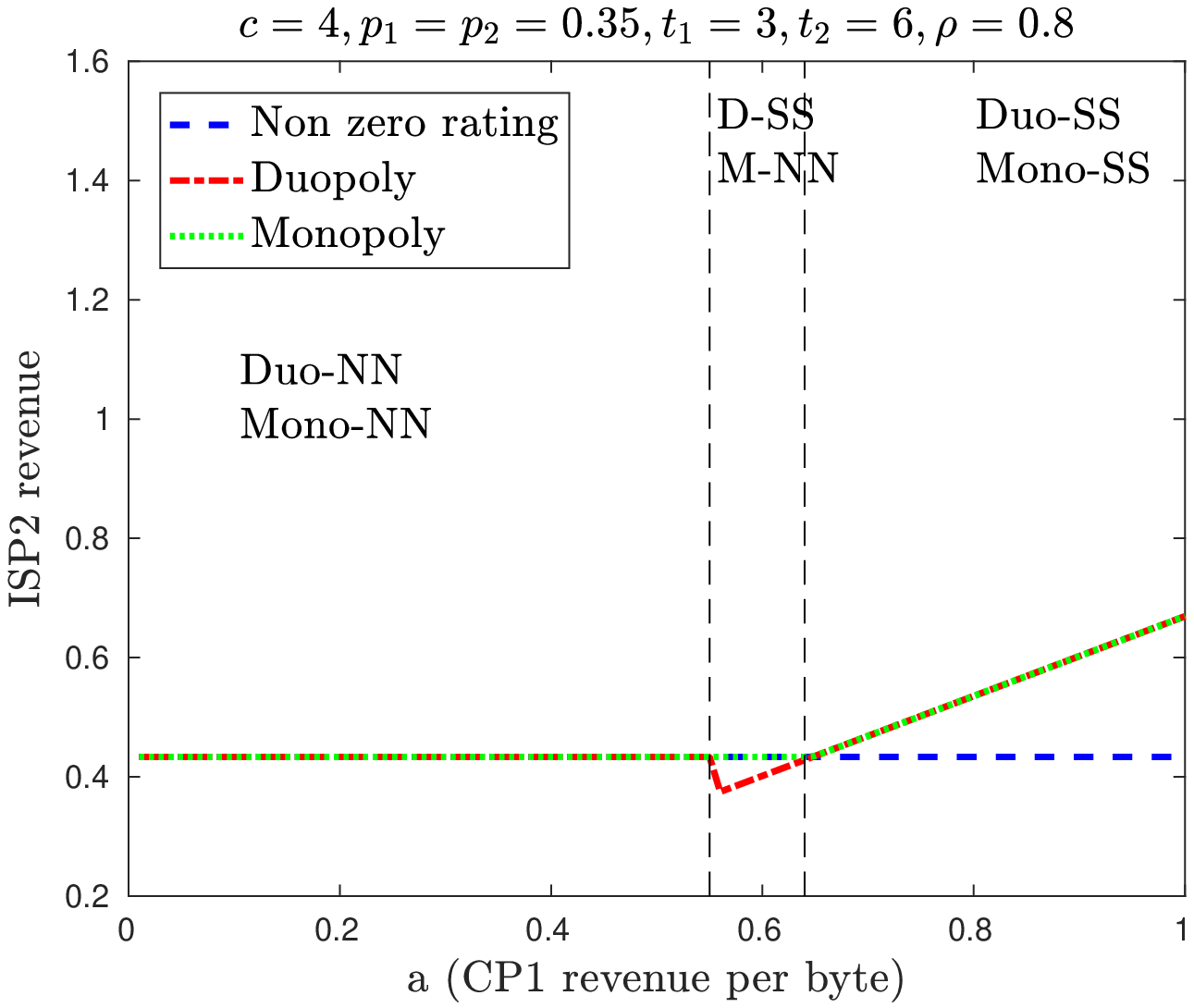}}
			\subfloat[]{
				\includegraphics{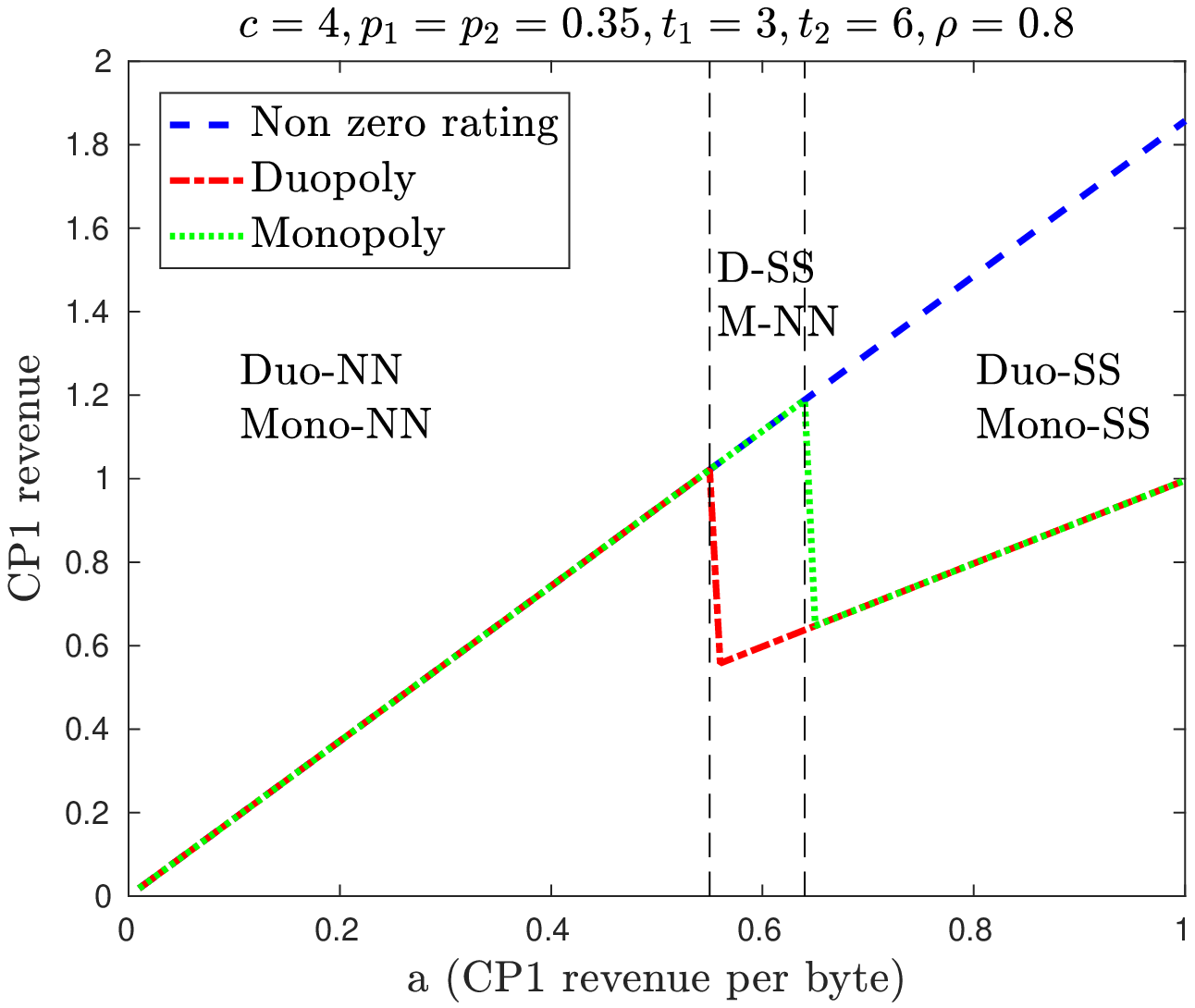}}
			\subfloat[]{
				\includegraphics{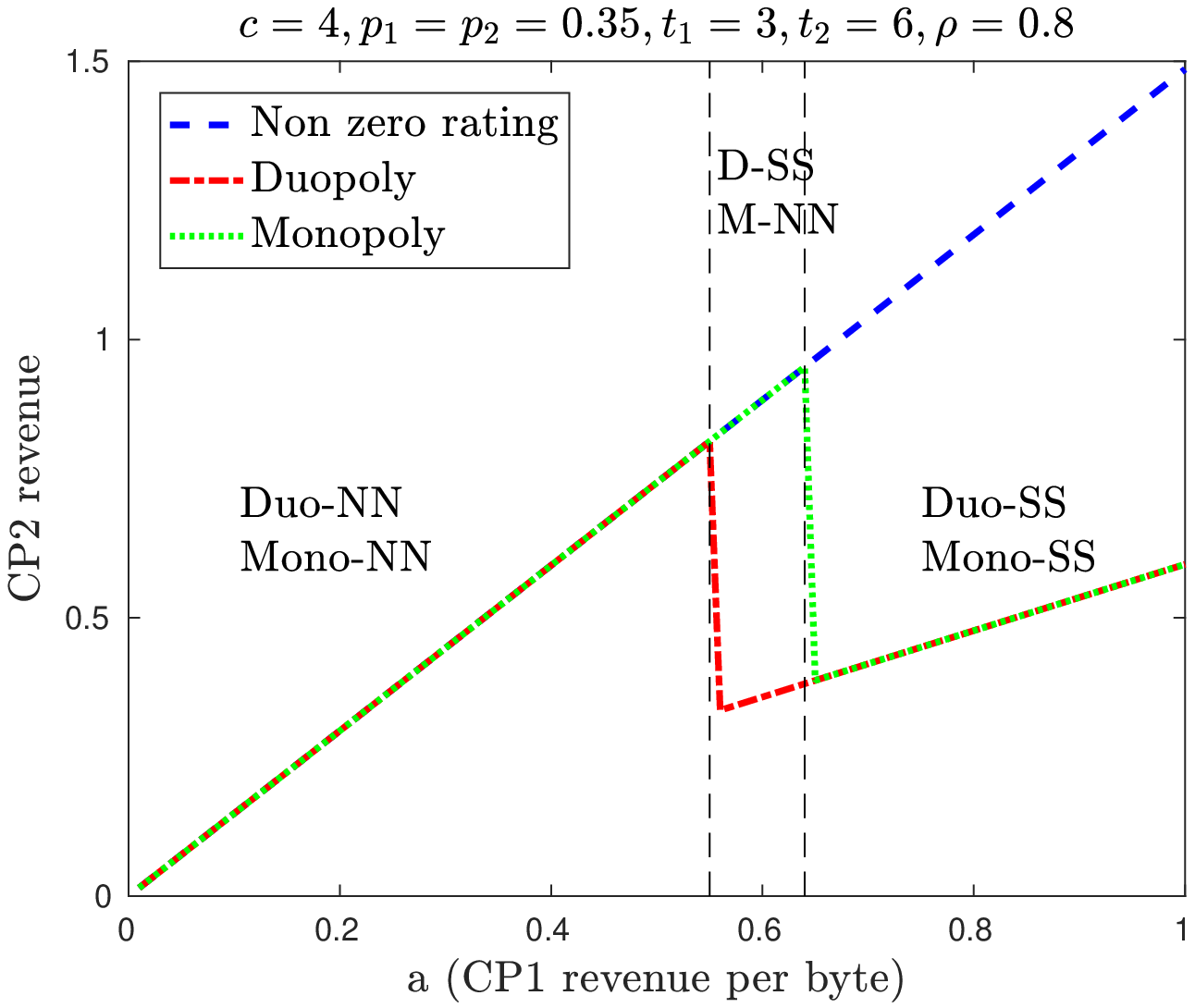}}

		}
		\caption{Surplus of various entities for $c=4,t_1=3,t_2=6,p_1=p_2=0.35$ and $\rho=0.8$ as a function of $a$ (a) ISP1 revenue (b) ISP2 revenue (c) CP1 revenue (d) CP2 revenue}
		\label{fig:assym_c4_p35_rho8}
	\end{center}
\end{figure*}

We simulate the alternating best response dynamics for this setting,
taking $t_1 = 3,$ and $t_2 = 6.$ As before,
$\psi(\theta) = \log(1 + \theta).$ Interestingly, the limiting
sponsorship configurations that emerge from best response dynamics
remain symmetric across the ISPs; see
Figure~\ref{fig:market_c90_p35}c. Moreover, the limiting configurations
are qualitatively identical to the case where user stickiness is
symmetric. Indeed, asymmetry in user-stickiness primarily manifests in
an asymmetry in the market shares of the two ISPs. To see this, we let
$(a_1,a_2) = (a, \rho a),$ and plot the equilibrium surpluses of the
ISPs, CPs, and the user base as a function of $a$ for $\rho = 0.1$
(see Figure~\ref{fig:assym_c4_p35_rho1}) and for $\rho = 0.8$ (see
Figure~\ref{fig:assym_c4_p35_rho8}). We benchmark the observed surplus
against the surplus when (i) neither ISP operates a zero-rating
platform, and (ii) the monopoly setting where each ISP's market share
is fixed to that under case (i). We observe that
\begin{enumerate}
\item As expected, ISP~1 enjoys a higher surplus than ISP~2, owing to
  its larger market share.
\item When $\rho$ is small, both ISPs induce an SN equilibrium for
  large enough $a.$
\item When $\rho$ is large, both ISPs induce an SS equilibrium for
  large enough $a.$
\item For intermediate values of $a,$ there is a prisoner's dilemma
  between the ISPs, where they both induce sponsorship prematurely,
  resulting in a lower surplus. Except in this region, the equilibrium
  sponsorship configurations and surpluses match those in the monopoly
  model.
\item At-least one CP, and sometimes both CPs, end up worse off due to
  zero-rating.
\end{enumerate}

\section{Policy Implications and Discussion}
\label{sec:discuss}

The key takeaway from our analyses is the following. When ISPs lead
in setting sponsorship prices, they do in such a way that a
significant fraction of the CP surplus gets paid to the ISPs in the
form of sponsorship costs. This reduces the CP surplus significantly.
Further, if one of the CPs is more profitable than the other, then
ISPs force a configuration in which the more profitable CP sponsors
and the other does not, skewing the consumption profile of the user
base.
This fact does not change with increased capacity of the users to
consume, or when the users are not sticky in their choice of
ISP. Therefore the ISPs have an interest in picking winners from among
the CPs. Interestingly, even the `winner' CP does not typically
benefit in this process! On the other hand, less profitable CPs can
suffer and be eliminated from the market.

In other words, data sponsorship practices grant ISPs considerable
market power---indeed, our analysis highlights that this power is
\emph{not} diminished by inter-ISP competition. Thus the meta message
from our analysis is that the zero rating, although good for the
consumers in the short term because of the increase in their surplus,
could in the long run have negative consequences on the CP
marketplace.

An important observation from our analysis is that zero rating drives
consumption away from non-sponsored content.\footnote{This has also
  been verified empirically. dflmonitor.eu has reported that the ISPs
  that provide zero rated content actually sell significantly less
  bandwidth to end users than those that do not zero-rate.} Indeed,
even when the CP profitability is small, ISP competition induces
sponsorship at smaller values of CP-profitability than in the monopoly
case. Since this skew of user consumption in favor of sponsored
content lies at the heart of the ISP market power, a possible
regulatory intervention (other than disallowing data sponsorship
entirely) could be to limit zero-rated content so as to leave room for
non zero-rated content to also contend for user attention.

It is important at this point to clarify the scope of our model and
our conclusions. Our leader-follower interaction model assumes the ISP
as the leader and the CPs as followers. This is natural when a `large'
ISP operates a zero-rating platform for `smaller' CPs. For example,
\textit{Sponsored Data} from AT\&T and \textit{FreeBee Data} from
Verizon.
However, it should be noted that there are also situations where the
dominance is reversed, e.g., the interaction between small ISPs and
large CPs like Google and Facebook. Such interactions are typically
based not on data sponsorship, but on peering arrangements, and would
require very different models. Early works on the economics of
Internet peering are \cite{Baake99,Cremer00} while
\cite{Courcoubetis16a,Courcoubetis16b,Ma17,Patchala19} are some recent
works analyzing paid peering.

\bibliographystyle{IEEEtran}
\bibliography{network-economics}

\appendices

\section{Proofs of results in Section~\ref{sec:model}}
\label{sec:app_NE-conditions}

\subsection{Proof of Lemma~\ref{lemma:theta_ordering}}
\label{sec:appendix:theta_ordering}

\begin{proof}
  Recall that user utility for both CPs is same, i.e.,
  $\psi_1(\cdot)=\psi_2(\cdot)=\psi(\cdot).$ From this it is easy to
  verify that $\theta_1^{NS} = \theta_2^{SN}.$ We will prove the lemma
  for $\theta_1^{NS}.$ It is easy to observe that $\theta_1^{NS}$ is
  the maximum solution of the following concave function over $[0,c]$:
  \begin{equation}
    f(x) = \psi(x) + \psi(c-x) -px \label{eq:theta_1_NS}
  \end{equation}
  We will prove the lemma by considering the following three
  cases:

  \textbf{Case 1: $\theta_1^{NS} < \theta_1^{SS}.$} Observe that
  $\theta_1^{SS} = c/2$ is the maximum of the concave function
  $g(x) = \psi(x) + \psi(c-x)$ over $[0,c].$ By \eqref{eq:theta_1_NS},
  $f(x) = g(x)- px.$ At $\theta_1^{SS},$
  $f'(\theta_1^{SS}) =g'(\theta_1^{SS}) -p = -p,$ which implies that
  $\theta_1^{NS} < \theta_1^{SS}.$
	
  \textbf{Case 2: $\theta_1^{NS} < \theta_1^{SN}.$} We will show that
  $\theta_1^{SN} > \theta_1^{SS}$ which combined with Case~1 statement
  will prove the result.  It is easy to observe that $\theta_1^{SN}$
  is the maximum solution of concave function
  $h(x) = \psi(x) + \psi(c-x) -p(c-x).$ Writing it in terms of
  optimization function for $\theta_1^{SS}$ we get
  $h(x) = g(x) -p(c-x).$ Thus $h'(\theta_1^{SS}) = p >0$ implying that
  $\theta_1^{SN} > \theta_1^{SS}.$
	
  \textbf{Case 3: $\theta_1^{NS} < \theta_1^{NN}.$} Note that
  $\theta_1^{NN}$ is the maximum solution of the concave function
  $d(x)= \psi(x) + \psi(c-x) -px -p(c-x).$ By \eqref{eq:theta_1_NS} we
  get $f(x)= d(x)+p(c-x)$ By similar arguments as Case~1,
  $\theta_1^{NS} < \theta_1^{NN}.$
\end{proof}
	
\subsection{Nash equilibrium between CPs}

We require the following notation.
{\small
\begin{align*}
  \alpha_1 &= 1- \frac{(x_{SN}^{M_2} - x_{NN}^{M_2}) \theta_{1}^{M_2} +
    x_{NN}^{M_2} \theta_{1}^{NN}}{x_{SN}^{M_2}
  \theta_{1}^{SN}} \\
  \alpha_2 &= \frac{(x_{SN}^{M_2} - x_{NN}^{M_2}) \theta_{1}^{M_2}}{x_{SN}^{M_2}
  \theta_{1}^{SN}} \mathbbm{1}_{\{(M_2=SN) || (M_2=SS)\}} \\
  \beta_1 &= 1- \frac{(x_{NS}^{M_2} - x_{NN}^{M_2}) \theta_{2}^{M_2} +
    x_{NN}^{M_2} \theta_{2}^{NN}}{x_{NS}^{M_2}
  \theta_{2}^{NS}} \\
  \beta_2 &= \frac{(x_{NS}^{M_2} - x_{NN}^{M_2}) \theta_{2}^{M_2}}{x_{NS}^{M_2}
  \theta_{2}^{NS}}\mathbbm{1}_{\{(M_2=NS) || (M_2=SS)\}} \\
  \gamma_1 &= 1- \frac{(x_{SS}^{M_2} - x_{NS}^{M_2}) \theta_{1}^{M_2} +
    x_{NS}^{M_2} \theta_{1}^{NS}}{x_{SS}^{M_2}
    \theta_{1}^{SS}} \\
\gamma_2 &= \frac{(x_{SS}^{M_2} - x_{NS}^{M_2}) \theta_{1}^{M_2}}{x_{SS}^{M_2}
           \theta_{1}^{SS}}\mathbbm{1}_{\{(M_2=SN) || (M_2=SS)\}} 
\end{align*}
\begin{align*}
  \delta_1 &= 1- \frac{(x_{SS}^{M_2} - x_{SN}^{M_2}) \theta_{2}^{M_2} +
             x_{SN}^{M_2} \theta_{2}^{SN}}{x_{SS}^{M_2}
  \theta_{2}^{SS}} \\
\delta_2 &= \frac{(x_{SS}^{M_2} - x_{SN}^{M_2}) \theta_{2}^{M_2}}{x_{SS}^{M_2}
	\theta_{2}^{SS}}\mathbbm{1}_{\{(M_2=NS) || (M_2=SS)\}} \\
\end{align*}
}
It is not hard to show the following.

\begin{lemma}
  \label{lemma:boundsCPNash}
  $\alpha_i, \beta_i, \gamma_i, \delta_i \in [0,1]$ for $i=1,2.$
\end{lemma}
We are now ready to state the conditions for each sponsorship
configuration on ISP~1 to be a Nash equilibrium.

\begin{lemma}
  Given a sponsorship configuration $M_2$ on ISP~2, the conditions for
  the different sponsorship configurations on ISP~1 to be Nash
  equilibrium between the CPs are:
  \begin{enumerate}
  \item NN is Nash equilibrium on ISP~1 if and only if
    \[ q_1 \geq \max(a_1 \alpha_1 +q_2\alpha_2, a_2 \beta_1+q_2\beta_2) \]
  \item SN is Nash equilibrium on ISP~1 if and only if
    \[ a_2 \delta_1 +q_2\delta_2 \leq q_1 \leq a_1 \alpha_1 + q_2\alpha_2 \]
  \item NS is Nash equilibrium on ISP~1 if and only if
    \[ a_1 \gamma_1 + q_2\gamma_2 \leq q_1 \leq a_2 \beta_1 + q_2\beta_2 \]
  \item SS is Nash equilibrium on ISP~1 if and only if
    \[ q_1 \leq \min(a_1 \gamma_1+q_2\gamma_2, a_2 \delta_1+q_2\delta_2) \]
  \end{enumerate}
  \label{lm:equilibrium_conditions}
\end{lemma}

 \section{Proof of Theorem~\ref{thm:opt_ISP1_market_str}}
 \label{sec:opt_ISP1_market_str} 

 \textbf{[Proof of Statement~1]} For fixed $a, \rho$ and $p$ ISP1 will
 set $q_1$ which maximizes its revenue. Let the maximum revenue of
 ISP1 in $M_1$ sponsorship configuration be $r_I^{M_1}(a).$ By
 Lemma~\ref{lm:equilibrium_conditions}, maximum revenue of ISP1 in
 various configurations is: \vspace{-0.1in}
  \begin{align}
    r_I^{NN}(a) &= x_{NN} p(\theta_{1}^{NN} + \theta_{2}^{NN}) \label{eq:isp_profit_nn} \\
    r_I^{SN}(a) &= x_{SN}p\theta_{2}^{SN} + x_{SN} a \alpha \theta_{1}^{SN} \label{eq:isp_profit_sn} \\
    r_I^{SS}(a) &= x_{SS} ac \min(\gamma,\rho \delta). \label{eq:isp_profit_ss}
  \end{align}
  Note that $r_I^{NN}(a)$ is constant with respect to $a$ while
  $r_I^{SN}(a)$ and $r_I^{SS}(a)$ are linearly increasing functions of
  $a.$ Thus there exists a $a=a'$ such that $r_I^{NN}(a') =
  r_I^{SN}(a').$ Similarly there exists $a=a''$ such that
  $r_I^{NN}(a'') = r_I^{SS}(a'').$ Therefore for $a < \min(a',a'')$ we
  get $\max(r_I^{SN}(a), r_I^{SS}(a)) \leq r_I^{NN}(a).$ Then we set
  $a_s = \min(a',a'')$ and for any $a\leq a_s$ ISP1 will enforce NN
  equilibrium by setting $q_1 \geq \max(a\alpha,a\rho \beta).$
	
  \textbf{[Proof of Statement~2]} For $a>a_s,$ ISP1 will maximize its
  revenue as: \vspace{-0.1in}
  \begin{align*}
    r_I(a) &= \max(r_I^{SN}(a), r_I^{SS}(a)) \\ &=
    \max(x_{SN}p\theta_{2}^{SN} + x_{SN} a \alpha
    \theta_{1}^{SN}, x_{SS} ac \min(\gamma,\rho \delta) ).
  \end{align*}
  As both the terms are increasing functions of $a,$ $r_I(a)$ is also
  an increasing function of $a.$ In this case, ISP1 will select SN
  over SS if $r_I^{SN}(a) \geq r_I^{SS}(a).$ In other words, if
  $x_{SN}p\theta_{2}^{SN} + x_{SN} a \alpha
  \theta_{1}^{SN} \geq x_{SS} ac \min(\gamma,\rho \delta)$ then
  ISP1 will set $q_1= a\alpha$ to get SN equilibrium else it will set
  $a_1=a\min(\gamma,\rho \delta)$ to get SS equilibrium.

\section{Proof of Lemma~\ref{lemma:opt_ISP1_threshold}}
\label{appendix:opt_ISP1_threshold}	

\begin{proof}
  When ISP2 is in NN state, for any transportation cost $t,$ the
  market share of ISP~1 in NN state is $x_{NN} =0.5$ for any $t.$
  Moreover, by Lemma~\ref{lemma:user_utility_sort}, $x_{SN}$ and
  $x_{SS}$ are both decreasing functions of $t.$ Recall the revenue of
  ISP~1 in NN, SN and SS state (see
  \eqref{eq:isp_profit_nn}-\eqref{eq:isp_profit_ss}).  Hence ISP~1
  revenue in NN state $r_I^{NN}$ is independent of $t$ while
  $r_I^{SN}$ and $r_I^{SS}$ are decreasing functions of $t.$ Thus $a'$
  such that $r_I^{NN}(a') = r_I^{SN}(a')$ is an increasing function of
  $t.$ Similarly $a''$ such that $r_I^{NN}(a'') = r_I^{SS}(a'')$ is an
  increasing function of $t.$ This shows that the threshold
  $a_S= \min(a',a'')$ is an increasing function of $t.$
\end{proof}

\section{Proof of Lemma~\ref{thm:thm:opt_ISP1_ISP1rev}}

\begin{proof}
  \textbf{[Proof of statement~1]} For $a\leq a_s,$ by
  Theorem~\ref{thm:isp_q1} ISP1 sets NN equilibrium by appropriately
  choosing $q_1.$ Thus ISP1's revenue in this case given by
  \eqref{eq:isp_profit_nn} which is constant with respect to $a.$

  \textbf{[Proof of statement~2]} For $a>a_s,$ by
  Theorem~\ref{thm:isp_q1} ISP1 sets SN or SS equilibrium and revenue
  in both equilibria is an increasing function of $a$ (see
  \eqref{eq:isp_profit_sn} and \eqref{eq:isp_profit_ss}).

\end{proof}

\section{Proof of Lemma~\ref{thm:cp_rev_only_p}}

\begin{proof}
  \textbf{[Proof of statement~1]} For $a>a_s,$ ISP1 sets $q=a\alpha$
  to get SN equilibrium. In this case profit of CP1 is:
  \[r_1^{SN} = x_{SN} (a-q_1) \theta_{1}^{SN} + (1-x_{SN}) 
  a \theta_{1}^{NN} \]
  By substituting value of $\alpha$ and
  simplifying we get,
  \begin{equation*}
  r_1^{SN} = a(x_{NN} \theta_{1}^{NN} +
  \theta{1}^{NN}(1-x_{NN})) =
  r_1^{NN}. 
  \end{equation*}
  Thus CP1
  profit is unaffected while going from NN to SN. CP2's profit when
  ISP1 is in SN configuration is:
  \[ r_2^{SN} = x_{SN} a \rho \theta_{2}^{SN} + (1-x_{SN}) 
  a \rho \theta_{2}^{NN}. \] 
  Similarly CP2's profit when ISP1 is in NN configuration is:
  \[ r_2^{NN} = x_{NN} a \rho \theta_{2}^{NN} + (1-x_{NN}) 
  a \rho \theta_{2}^{NN}.\]
  By subtracting $r_2^{NN}$ from $r_2^{SN}$ and simplifying we get,
  \begin{equation*}
    r_2^{SN} - r_2^{NN} = a\rho \left[x_{SN} 
      (\theta_{2}^{SN} - \theta_{2}^{NN}) \right] \leq 0.
  \end{equation*}
  The last inequality follows from Lemma~\ref{lemma:theta_ordering}. Thus,
  \begin{equation}
    r_2^{SN} \leq r_2^{NN} \label{eq:cp2_profit_sn_nn}
  \end{equation}
  This proves that CP2 is worse off in SN configuration of ISP1
  compared to non-zero rating setting.
  
  \textbf{[Proof of statement~2]} By
  Lemma~\ref{lm:equilibrium_conditions} to get SS configuration, ISP1
  sets $q_1 = a \min(\gamma, \rho \delta).$ Thus we will analyze
  profits of CPs in the following two cases:
  \begin{enumerate}
  \item For $q_1=a\rho \delta,$ CP2 profit is:
  \begin{align*}
  r_2^{SS} &= x_{SS}(\rho a - \delta \rho a) \theta_{2}^{SN}
  + (1-x_{SS}) \rho a \theta_{2}^{NN} \nonumber \\ &= a \rho
  (x_{SN} \theta_{2}^{SN}+ (1-x_{SN})
  \theta_{2}^{NN}) \nonumber \\ &= r_2^{SN}. 
  \end{align*}

    By \eqref{eq:cp2_profit_sn_nn} we know
    $r_2^{SN} \leq r_2^{NN}.$ Thus, $r_2^{SS} \leq r_2^{NN}$ making
    CP2 worse-off than in NN configuration.
  \item For $q_1 = a\gamma$ it is easy to prove that $r_1^{SS} =
    r_1^{NS}.$ Now we will prove that $r_1^{NS} \leq r_1^{NN}.$ We
    know that
    \[r_1^{NS} = x_{NS} a \theta_{1}^{NS} + (1-x_{NS}) 
    a \theta_{1}^{NN}, \]
    and
    \[r_1^{NN} = x_{NN} a \theta_{1}^{NN} + (1-x_{NN}) 
    a \theta_{1}^{NN}. \]
    By subtracting $r_1^{NN}$ from $r_1^{NS}$ we get,
    \begin{equation*}
      r_1^{NS} - r_1^{NN} = a x_{NS}(\theta_1^{NS}-\theta_1^{NN}) \leq 0.
    \end{equation*}
    Here last inequality follows from Lemma~\ref{lemma:theta_ordering}. Thus
    $r_1^{SS} \leq r_1^{NN}$ which shows that CP1 is worse-off than in
    non zero rating setting in this case.
  \end{enumerate}
 
\end{proof}

\section{Proof of Theorem~\ref{thm:symmetric}}
\label{sec:sym_equilibrium_cp_proof}

      Consider first the case $M = SN;$ the case $M = NS$ follows via
      a symmetric argument. If $SN-SN$ is a system equilibrium, it can
      be shown that $a_1 \geq a_2;$ indeed, if not, any ISP has the
      incentive to switch to an NS configuration. From the proof of
      Theorem~\ref{thm:isp_eq_a_large_rho_small}, it then follows that
      under an SN-SN equilibrium, both ISPs would set their
      sponsorship price as
      $q=a_1 \left(1- \frac{\theta_{1}^{NN}}{\theta_{1}^{SN}}
      \right).$

      The profit of the sponsoring CP (CP~1) in this case equals
      $x_{SN}^{SN} \theta_1^{SN} (a-q) +
      (1-x_{SN}^{SN})\theta_1^{SN}(a_1-q) = a_1 \theta_1^{NN},$ which
      equals its profit in the absence of zero-rating. Now, profit of
      the non-sponsoring CP (CP~2) under this configuration equals
      $x_{SN}^{SN} a_2 \theta_2^{SN} + (1-x_{SN}^{SN}) a_2 
      \theta_2^{SN} = a_2 \theta_2^{SN} < a_2  \theta_2^{NN}.$
      Thus, CP~2 is worse off under this configuration, relative to
      the scenario where zero-rating is not permitted.
	
      Next, we now consider $M = SS.$ Assume WLOG that $a_1 \geq a_2.$
      It is easy to show that under a symmetric equilibrium, both ISPs
      would set $q = a_2 \left(1- \frac{\theta_2^{SN}}{c/2}\right).$
      The revenue of CP2 in this configuration is
      $x_{SS}^{SS} (a\rho-q) \theta_2^{SS} + (1-x_{SS}^{SS}) (a \rho
      -q) \theta_2^{SS} = a \rho \theta_2^{SN} < a_2 \theta_2^{NN}.$
      Thus, when ISPs are in symmetric SS-SS configuration, CP2 is
      worse off relative to the setting where zero-rating is not
      permitted.

\section{Proof of Theorem~\ref{thm:isp_eq_a_large_rho_small}}
\label{sec:a_large_rho_small_proof}

Suppose that ISP~1 and ISP~2 have both induced an SN state with equal
$q=q_1=q_2.$ By Lemma~\ref{lm:equilibrium_conditions}, to maximize
profit in SN state, ISP~1 would set $q_1=a\alpha_1 +q_2\alpha_2.$
Similarly ISP~2 would set $q_2=a\alpha_1 +q_1\alpha_2.$ For
$q=q_1=q_2$ we get, $q=\frac{a\alpha_1}{1-\alpha_2}.$

  Substituting $x_{SN}^{SN} =0.5$ in the expressions for
  $\alpha_1,\alpha_2$ assuming other ISP is in SN state we get,
  \begin{align*}
    \alpha_1 = \frac{x_{NN}^{SN} (\theta_{1}^{SN} - \theta_{1}^{NN})}{0.5 \theta_{1}^{SN}}, \quad \alpha_2 = \frac{0.5-x_{NN}^{SN}}{0.5}.
  \end{align*}
  Substituting these values in expression for $q$ we get,
  \begin{equation*} 
    q = q(a) := a \left(1- \frac{\theta_{1}^{NN}}{\theta_{1}^{SN}} \right). \label{eq:q_sn_sn}
  \end{equation*}

  We now show that $(q(a),\text{SN},q(a),\text{SN})$ is a system
  equilibrium. The revenue of ISP~1 in this configuration is
  \begin{equation*}
	   r_1^{SN} = x_{SN}^{SN}(q \theta_{1}^{SN} + p \theta_{2}^{SN}) 
	\end{equation*}
	Noting that $\theta_{2}^{SN} = c-\theta_{1}^{SN}$ and
        substituting the value of $q$ in above expression we get,

        $r_1^{SN} = 0.5(a(\theta_{1}^{SN} - \theta_{1}^{NN}) +
        p(c-\theta_{1}^{SN})).$
         
	By Lemma~\ref{lemma:theta_ordering}, for $a > \frac{p
          \theta_{1}^{SN}}{(\theta_{1}^{SN} - \theta_{1}^{NN})} =
        a_{SN},$ $r_1^{SN} > 0.5pc.$ Now we will show that ISP~1
        cannot increase its profit by switching to a different
        sponsorship configuration.  For this we need to show that the
        if ISP~1 moves to NN or SS configuration given ISP~2 is in SN,
        then ISP~1's revenue will not increase from $r_1^{SN}.$ Let
        $\hat{r}_1^{NN}$ be ISP~1's revenue if it moves to NN
        configuration given ISP~2 is in SN. Then,
	\begin{equation*}
	   \hat{r}_1^{NN} = x_{NN}^{SN}p(\theta_{1}^{NN} + \theta_{2}^{NN}) < 0.5 pc,
	\end{equation*}
	where the last inequality follows from $x_{NN}^{SN} \leq
        x_{SN}^{SN} = 0.5$ and $\theta_{1}^{NN} + \theta_{2}^{NN}\leq
        c.$ Thus for $a> a_s,$ $r_1^{SN} > \hat{r}_1^{NN}.$ So, ISP~1
        will not switch to an NN configuration.
	
	Let $\hat{r}_1^{SS}$ be the ISP~1 revenue if it moves to SS
        configuration given that ISP~2 is in the SN state. Then, we
        know that
        $\hat{r}^{SS} = x_{SS}^{SN} c \min(a\gamma_1+\gamma_2, a \rho
        \delta_1).$ Note that $\delta_2=0$ as ISP2 is in SN state. For
        small enough $\rho,$
        $\hat{r}^{SS} = x_{SS}^{SN} c a \rho \delta_1.$ Substituting
        the appropriate terms for $\delta_1$ we get,
        $\hat{r}_1^{SS} = x_{SS}^{SN}c a \rho \left(1- \frac{2
            \theta_{22}^{SN}}{c} \right).$
	As $\theta_{2}^{SN} \in [0,c/2)$ we get
        $\hat{r}_1^{SS} \leq x_{SS}^{SN} a \rho c.$ Also
        $c-\theta_{1}^{SN} \geq 0$ implying
        $r_1^{SN} \geq 0.5(a (\theta_{1}^{SN} - \theta_{1}^{NN}))$
        thus for
	\begin{equation*}
	  \rho < \frac{0.5 (\theta_{1}^{SN} - \theta_{1}^{NN})}{x_{SS}^{SN} c} = \rho_{SN}
	\end{equation*}
	we get $r_1^{SN} > \hat{r}_1^{SS}.$ 

        This proves that $(q(a),\text{SN},q(a),\text{SN})$ is a system
        equilibrium. It is not hard to show that $(q(a),\text{SN})$ is
        the optimal configuration for each ISP in the monopoly
        setting.

        Finally, we note that if any CP is to switch its sponsorship
        decision on both ISPs, the market split would remain
        equal. Therefore, that SN is a Nash equilibrium between CPs in
        the monopoly setting implies that neither CP has the incentive
        to switch its sponsorship decision on both ISPs.

      \section{Proof of Theorem~\ref{thm:isp_eq_a_large_rho_large}}
      \label{sec:a_large_rho_large_proof}
      \begin{proof}
        Suppose that ISP~1 and ISP~2 both induce an SS configuration
        with $q_1=q_2=q.$ Then by
        Lemma~\ref{lm:equilibrium_conditions},
        $q_1 \leq \min(a\gamma_1+q_2\gamma_2,
        a\rho\delta_1+q_2\delta_2).$ First we will prove that
        $q_1= a\rho\delta_1+q_2\delta_2$ when both ISPs are in SS
        state. Recall the expressions for $\gamma_1$ and $\delta_1:$
   \begin{align*}
      \gamma_1 &= 1- \frac{(x_{SS}^{SS}-x_{NS}^{SS})\theta_{1}^{SS} + x_{NS}^{SS} \theta_{1}^{NS}}{x_{SS}^{SS}\theta_{1}^{SS}} \\
      \delta_1 &= 1- \frac{(x_{SS}^{SS}-x_{SN}^{SS})\theta_{2}^{SS} + x_{SN}^{SS} \theta_{2}^{SN}}{x_{SS}^{SS}\theta_{2}^{SS}}
   \end{align*}
   As utility derived from both CPs is same for a user,
   $x_{NS}^{SS} = x_{SN}^{SS},$ and
   $\theta_{1}^{SS} = \theta_{2}^{SS}=c/2,$ thus $\gamma_1=\delta_1.$
   Similarly, $\gamma_2=\delta_2.$ As $\rho < 1,$ we get
   $q_1= a\rho \delta_1 + q_2 \delta_2.$ Simplifying this equation for
   $q=q_1=q_2$ we get
   \begin{equation}
     q = q(a,\rho) := a \rho \left( 1- \frac{\theta_{2}^{SN}}{c/2} \right). \label{eq:q_ss_ss}
   \end{equation} 
   Revenue of ISP1 in this state is $r_1^{SS} = x_{SS}^{SS} c q = 0.5 c a \rho \left( 1- \frac{\theta_{2}^{SN}}{c/2} \right)$ which is a function of $a$ and $\rho.$ Now we will prove that ISP1 will not move to NN or SN state from here so as to increase its revenue. Let $\hat{r}_1^{NN}$ be ISP1's revenue when ISP2 is in SS which can be written as $\hat{r}_1^{NN} = x_{NN}^{SS}p (\theta_{1}^{NN}+ \theta_{1}^{NN}).$ Thus $\hat{r}_1^{NN}$ is independent of $a$ or $\rho.$ Thus there exists $a>a_n$ and $\rho >\rho_n$ such that $\hat{r}_1^{NN} < r_1^{SS}$ in which case ISP1 will not move to NN from SS. 
   
   Let $\hat{r}_1^{SN}$ be the ISP1 revenue if it moves to SN given
   ISP2 is in SS. Then,
   $\hat{r}_1^{SN} = x_{SN}^{SS}(q_{SN}\theta_{1}^{SN}+ p
   \theta_{2}^{SN})$ where $q_{SN}$ is value of $q_1$ for SN state
   given ISP2 is in SS. By Lemma~\ref{lm:equilibrium_conditions}
   $q_{SN} = a \alpha_1 +q \alpha_2$ where $q$ is given by
   \eqref{eq:q_ss_ss}. Substituting value of $q_{SN}$ and simplifying
   we get,
   \begin{align*}
      \hat{r}_1^{SN} &= x_{SN}^{SS}(c(a\alpha_1+q\alpha_2) + \theta_{2}^{SN}(p-a \alpha_1-q\alpha_2)) \\
      &< 0.5(c(a\alpha_1+q\alpha_2)) \\
      &= 0.5 ac \left(\alpha_1 + \rho \alpha_2 \left( 1- \frac{\theta_{2}^{SN}}{c/2} \right)\right),
   \end{align*}
   where inequality holds for any $a>p/\alpha_1 = a_s.$ Comparing this with $r_1^{SS}$ we can say that if
   \begin{equation*}
      \rho > \frac{\alpha_1}{1-\alpha_2} \frac{1}{1- \frac{\theta_{2}^{SN}}{c/2}} = \rho_n
   \end{equation*}
   then $\hat{r}_1^{SN} < r_1^{SS}.$ Thus if we choose
   $a_{SS} =\max(a_s,a_n)$ and $\rho_{SS}=\max(\rho_s,\rho_n)$ then
   ISP1 will not move away from SS configuration.

   The rest of the argument is identical to that in the proof of
   Theorem~\ref{thm:isp_eq_a_large_rho_small}.
\end{proof}

\end{document}